\def\reporttype{1}
\def\PLAINreport{0}
\def\INRIAreport{1}
\ifnum\reporttype<0\stop\fi
\ifnum\reporttype>1\stop\fi

\documentclass[11pt]{article}

\usepackage{url}
\usepackage{makeidx}
\usepackage{multicol}
\usepackage{amsmath,amssymb}
\usepackage[dvips]{graphicx}
\usepackage{psfrag}
\usepackage{color}

\ifnum\reporttype=\INRIAreport
\usepackage{RR}
\fi

\ifnum\reporttype=\INRIAreport
\else
\fi

\def\withrefnotes{0}
\def\withaddnotes{0}
\def\withffootnotes{1}
\def\withenglishnotes{0}

\ifnum\reporttype=\PLAINreport\else
\def\withaddnotes{0}
\def\withffootnotes{0}

\def\withenglishnotes{0}
\fi

\newcommand{\addnote}[1]{\ifnum\withaddnotes=0\else{\footnote{#1}}\fi}
\newcommand{\englishnote}[1]{\ifnum\withenglishnotes=0%
\else{\footnote{{\em English\/}: #1}}\fi}
\newcommand{\ffootnote}[1]{%
\ifnum\withffootnotes=0\else{\footnote{#1}}\fi}
\newcommand{\refnote}[1]{\ifnum\withrefnotes=0\else{\footnote{#1}}\fi}

\ifnum\withrefnotes=1
\usepackage[first,bottomafter]{draftcopy}
\draftcopyName{Refereeing}{150}
\fi

\numberwithin{equation}{section}

\newcommand\qedsymbol{\hbox{\rlap{$\sqcap$}$\sqcup$}}
{\noindent{\sc Proof}.}%
{\qed}

\if{

}\fi

\newcommand{\COLL}{\textit{\texttt{coll}}}
\newcommand{\COMP}{\textit{\texttt{comp}}}

\newcommand{\MACH}{\textit{\texttt{mach}}}
\newcommand{\OS}{\textit{\texttt{os}}}
\newcommand{\PROB}{\textit{\texttt{prob}}}

\newcommand{\SOLV}{\textit{\texttt{solv}}}
\newcommand{\SOLVa}{\textit{\texttt{solv1}}}
\newcommand{\SOLVb}{\textit{\texttt{solv2}}}

\newcommand{\SUBC}{\textit{\texttt{subc}}}
\newcommand{\TAG}{\textit{\texttt{tag}}}

\newcommand{\abr}{\allowbreak}
\newcommand{\bulindent}{1.3\parindent}

\if{
\def\<#1,#2>{\langle#1,#2\rangle}

\newcommand{\gbar}[1]{\newbox\charbox\setbox\charbox=\hbox{$#1$}%
\vbox{\vbox{\hrule width\wd\charbox height0.3pt}\nointerlineskip%
\kern0.12em\box\charbox}}

\def\mat#1#2{%
  \if#1<%
    \if#2=\preccurlyeq\else\prec#2\fi%
  \else\if#1>%
         \if#2=\succcurlyeq\else\succ#2\fi%
       \else#1#2%
       \fi%
  \fi}

\newcommand{\qed}%
{\hspace*{1em}\hfill\qedsymbol}

}\fi

\newcommand{\theabstract}{%
The Libopt environment is both a methodology and a set of tools that
can be used for testing, comparing, and profiling solvers on problems
belonging to various collections. These collections can be
heterogeneous in the sense that their problems can have common features
that differ from one collection to the other. Libopt brings a unified
view on this composite world by offering, for example, the possibility
to run any solver on any problem compatible with it, using the same
Unix/Linux command. The environment also provides tools for comparing
the results obtained by solvers on a specified set of problems. Most of
the scripts going with the Libopt environment have been written in
Perl.
}

\ifnum\reporttype=\INRIAreport\relax
\RRtitle{%
LIBOPT -- Un environnement pour \'evaluer des solveurs sur des
collections h\'et\'erog\`enes de probl\`emes\\-- Version 1.0 --
}
\RRetitle{%
LIBOPT -- An environment for testing solvers on heterogeneous
collections of problems\\-- Version 1.0 --
}
\titlehead{The Libopt environment}
\RRauthor{%
  J.\ Charles {\sc Gilbert}\/%
  \thanks{INRIA Rocquencourt, projet Estime, BP~105, 78153~Le Chesnay
  Cedex, France\,;
  e-mail\,: {\tt Jean}\abr{\tt -}\abr{\tt Charles.}\abr%
            {\tt Gilbert@}\abr{\tt inria.fr}.}
\and
  Xavier {\sc Jonsson}\/%
  \thanks{Mentor Graphics (Ireland) Ltd. - French Branch; 180, Avenue
  de l'Europe - Zirst Montbonnot; F-38334 Saint Ismier Cedex; e-mail:
  \texttt{Xavier\_Jonsson@mentorg.com}.}
}
\authorhead{J.\ Ch.\ Gilbert, X. Jonsson}
\RRdate{22 f\'evrier 2007}
\RRresume{%
L'environnement Libopt est \`a la fois une m\'ethodologie et un
ensemble d'outils qui peuvent \^etre utilis\'es pour tester, comparer
et profiler des solveurs sur des probl\`emes de diverses collections.
Ces derni\`eres peuvent \^etre h\'et\'erog\`enes dans le sens o\`u
leurs probl\`emes peuvent avoir des caract\'eristiques communes qui
diff\`erent d'une collection \`a l'autre. Libopt apporte un point de
vue unificateur sur ce monde composite en offrant, par exemple, la
possibilit\'e de lancer n'importe quel solveur sur n'importe quel
probl\`eme compatible avec lui, en utilisant la m\^eme commande
Unix/Linux. L'en\-vi\-ron\-ne\-ment fournit \'egalement des outils pour
comparer les r\'esultats obtenus par divers solveurs sur un jeu
sp\'ecifi\'e de probl\`emes. La plupart des scripts qui accompagnent
l'environnement Libopt ont \'et\'e \'ecrits en Perl.
}
\RRabstract{\theabstract}
\RRmotcle{%
calcul scientifique --
collection de probl\`emes --
comparaison de solveurs --
CUTEr --
environnement de test --
\'evaluation de performance --
Libopt --
Modulopt --
optimisation --
profil de performance
}
\RRkeyword{%
benchmarking --
collection of problems --
CUTEr --
Libopt --
Modulopt --
optimization --
performance profile --
scientific computing --
solver comparison --
testing environment.
}
\RRprojet{Estime}
\RRtheme{\THNum}
\URRocq
\fi

\makeindex

\makeatletter
  {\end{multicols}}
\makeatother

\index{+@\texttt{+}|see{regular expression}}
\index{CUTEr|see{collection}}
\index{Libopt line!tag in the --|see{tag}}
\index{Modulopt|see{collection}}
\index{addopt@\texttt{addopt}|see{command, file}}
\index{all.lst@\texttt{all.lst}|see{list}}
\index{blank \u@blank \texttt{\char`\\u}|see{regular expression}}
\index{coll@\COLL|see{collection}}
\index{comment|see{Libopt line, list}}
\index{default.lst@\texttt{default.lst}|see{list}}
\index{command (Libopt --)!perfopt@\texttt{perfopt}|seealso{directive}}
\index{dtbopt@\texttt{dtbopt}|see{file}}
\index{echo@\texttt{echo}|see{command}}
\index{grep@\texttt{grep}|see{command}}
\index{iflnx@\texttt{iflnx}|see{platform}}
\index{info@\texttt{info}|see{token}}
\index{install libopt@\texttt{install\_libopt}|see{command}}
\index{LIBOPT\_DIR@\texttt{LIBOPT\_DIR}|see{environment variable}}
\index{LIBOPT\_PLAT@\texttt{LIBOPT\_PLAT}|see{environment variable}}
\index{liboptrc@\texttt{liboptrc}|see{file}}
\index{file|seealso{list}}
\index{MASTSIF@\texttt{MASTSIF}|see{environment variable}}
\index{name|see{collection, problem, solver, subcollection}}
\index{perfopt@\texttt{perfopt}|see{command, file}}
\index{performance!token|see{token}}
\index{prob@\PROB|see{problem}}
\index{prompt|see{command prompt}}
\index{runopt@\texttt{runopt}|see{command}}
\index{s@\texttt{\char`\\s}|see{regular expression}}
\index{solv@\SOLV|see{solver}}
\index{solv\_coll@\texttt{\SOLV\_}\abr \texttt{\COLL}|see{command}}
\index{solver!tag|see{tag}}
\index{space \s@space \texttt{\char`\\s}|see{regular expression}}
\index{suffix!.lst@\texttt{.lst}|seealso{list}}
\index{suffix!.spc@\texttt{.spc}|see{file}}
\index{u@\texttt{\char`\\u}|see{regular expression}}
\index{w@\texttt{\char`\\w}|see{regular expression}}

\begin{document}


\ifnum\reporttype=\INRIAreport\else
\pagestyle{plain}

\renewcommand{\thefootnote}{\fnsymbol{footnote}}
\setcounter{footnote}{1}

\begin{center}\Large\bf
LIBOPT -- An environment for testing
solvers on heterogeneous collections of problems\\
\vspace{2ex}\rm\normalsize
Version 1.0~~~(\today)\\
\vspace{2ex}\rm\normalsize
J.\ Charles {\sc Gilbert}\/\footnote{INRIA-Rocquencourt, BP~105,
F-78153~Le Chesnay Cedex, France; e-mail:
\url{Jean-Charles.Gilbert@inria.fr}.}
and X. {\sc Jonsson}\/\footnote{Mentor Graphics (Ireland) Ltd. - French
Branch; 180, Avenue de l'Europe - Zirst Montbonnot; F-38334 Saint
Ismier Cedex; e-mail: \url{Xavier_Jonsson@mentorg.com}.}
\end{center}

\begin{abstract}
\noindent
\theabstract
\end{abstract}
\fi


\ifnum\reporttype=\INRIAreport\relax
\makeRT
\fi

\renewcommand{\thefootnote}{\arabic{footnote}}
\setcounter{footnote}{0}

\section{Introduction}
\label{s:introduction}
\ifnum\reporttype=\INRIAreport\else
\thispagestyle{empty}
\fi

Two of the issues that come up with software development have to do
with benchmarking and profiling solvers on some (possibly large)
collections of problems. For example, in the optimization community,
these issues are frequently dealt with the {\em CUTEr testing
environment\/}\index{collection!CUTEr}~\cite{bongartz-conn-gould-toint-1995,
gould-orban-toint-2003}. This one supplies ready-to-use interfaces
between some known solvers and a collection of problems encoded with
the SIF language. It is intended to help developers to test their
optimization code. The Libopt package has been designed for a similar
purpose but, unlike CUTEr, it provides neither collections of problems
written in a specific format, nor decoders for converting problems
written in some language into {\sc Fortran} or~C; on the other hand, it
is not restricted to optimization. Rather, Libopt has been thought up
for organizing and using problems coming from heterogeneous sources.
Heterogeneity refers, in particular, to the variety of languages used
to write the problems. As a result, CUTEr is considered by Libopt as a
particular collection of problems having its own features and
solver-collection interface.

Originally, we had in mind to make the problems of the {\em Modulopt
collection\/}\index{collection!Modulopt}~\cite{lemarechal-1980} as
easily available as the CUTEr problems, despite the diversity of their
encoding. The Modulopt collection is formed of problems coming from
industrial or scientific computing sources. Because of their
complexity, these codes are often not modeled on the academic standard
of idealized problems, so that it would have been difficult and boring
to rewrite them in SIF. This state of affairs motivated the development
of the Libopt environment, which resulted in a layer covering various
collections of problems (including CUTEr and Modulopt) and solvers,
organizing and normalizing the communication between them.

In addition to its solver-collection setting, Libopt also provides a
number of programs, mainly Perl scripts, that perform repetitive tasks,
such as running solvers on problems, collecting their results, and
comparing them. The features of the solvers and collections are
actually encoded in these programs, some of them having to be written
for each solver-collection pair. Once these are available, all the
results are obtained and compared using the same Unix/\abr Linux
commands. This is one of the interest of the proposed approach. Now,
Libopt is an open structure, which can grow, depending on the needs of
its users, by incorporating more solvers, problems, or even collections
of problems. A great part of this paper describes this aspect of the
software.


The suffix ``opt'' of the environment name, and of its various
companion scripts and programs, reveals the fact that Libopt was first
introduced and used to deal with optimization solvers and problems.
However, the concepts implemented in this software are sufficiently
general for being suitable to other scientific computation fields. The
tools have been designed with this idea of generality in mind.

The paper is organized as follows. The guided tour of
section~\ref{s:tour} invites the reader to discover the main aspects of
the software. This view of Libopt, from 10,000 feet, essentially
presents the tools that are intended to be routinely used.
Section~\ref{s:install} describes the Libopt package and its directory
structure; it enumerates the installation instructions. A~deeper
understanding of the Libopt mechanisms is necessary if one wishes to
enrich the environment with new collections of problems and solvers;
these ``squalid details'' are presented in section~\ref{s:depths}. We
consider in sections~\ref{s:adding-collection} and \ref{s:solver}
respectively how to install a new collection of problems and a new
solver. The paper ends with a prospective section that presents
possible improvements.

\paragraph{Notation.}
We use the following typographic conventions. The \texttt{typewriter
font} is used for a text that has to be typed literally and for the
name of files and directories that exist as such (without making
substitutions). In the same circumstances, a generic word, which has to
be substituted by an actual word depending on the context, is written in
\texttt{\textit{italic typewriter font}}.

Throughout, we assume that the operating system is Unix or Linux. The
{\em command prompt\/}\index{command prompt@command prompt
(\texttt{\%})} is denoted by the character `\texttt{\%}'. Optional
arguments in a command line are surrounded by the brackets `\texttt{[}'
and `\texttt{]}'.

We use the following abbreviations of regular expressions:
\begin{quote}
\texttt{\char`\\u}\index{regular expression!\u@\texttt{\char`\\u}
(blank character)} for \texttt{[\char`\ \char`\\t]} (the {\em
blank\/} character),\\
\texttt{\char`\\s}\index{regular expression!\s@\texttt{\char`\\s}
(space character)} for
\texttt{[\char`\ \char`\\f\char`\\n\char`\\r\char`\\t]} (the {\em
space\/} character),\\
\texttt{\char`\\w}\index{regular expression!\w@\texttt{\char`\\w}
(character for writing words)} for \texttt{[a-zA-Z0-9\_]} (the
character for writing words),
\end{quote}
where \texttt{\char`\\f} stands for a formfeed,
\texttt{\char`\\n} for a newline,
\texttt{\char`\\r} for a carriage return, and
\texttt{\char`\\t} for a tab.

\section{A guided tour}
\label{s:tour}

The Libopt environment has been designed to make easier and faster the
repetitive tasks linked to testing, comparing, and profiling solvers on
problems coming from heterogeneous collections of problems. These tasks
can be divided into three stages: running solvers on problems,
gathering the results, and comparing them. Libopt has three commands
(and some others) associated with these stages: \texttt{runopt},
\texttt{addopt}, and \texttt{perfopt}. This section offers an
introduction to them. Note that manual pages are available, which
give the details on the use of each of these commands. But before this,
let us start by defining terms that are continually used in this paper.

\subsection{A few definitions}

The {\em Libopt hierarchy\/}\index{Libopt hierarchy} is the set of
directories and files that form the Libopt environment. The directory
from which the Libopt scripts are launched is called below the {\em
working directory}\index{working directory|ibf}. The scripts take care
that this directory is not in the Libopt hierarchy. If this were the
case, there would be a danger of incurable destruction. Indeed, a
script like \texttt{runopt}\index{command (Libopt
--)!runopt@\texttt{runopt}} generally removes several files from the
working directory after a problem has been solved.

A {\em solver\/}\index{solver} is a computer program that can find the
solution to some classes of problems. Hem! Not sure this is very
explicit, but we shall not try to be more precise; in particular, we
shall take the notion of {\em problem\/}\index{problem} as a premiss,
not requiring any definition. The {\em name\/}\index{solver!name of a
--} of a solver is a string that must be match by \texttt{\char`\\w+}
(see the notation above for the meaning of the character
`\texttt{\char`\\w}'; the multiplier `\texttt{+}'\index{regular
expression!+@\texttt{+} (multiplier, at least once)} means that one or
more of the immediately previous character or character class, here
`\texttt{\char`\\w}', must be present).

A {\em collection\/}\index{collection} is a set of problems in an
arbitrary scientific computation area. The {\em
name\/}\index{collection!name of a --} of a collection is also a string
that must be match by \texttt{\char`\\w+}. Problems in the same
collection must differ by their {\em name\/}\index{problem!name of a
--} (again a string that must be matched by \texttt{\char`\\w+}), but
two problems belonging to two different collections may have the same
name. There are good reasons for gathering problems having common
features in the same collection. For instance, problems in a collection
are often written in the same language (Fortran, C, Matlab, Scilab,
Ampl, Gams, SIF, to mention a few). The motive is that, in the Libopt
environment, all the problems of a collection are dealt with the same
scripts and that these scripts are easier to conceive if the problems
are written in the same language. For the same reason, problems in a
collection usually belong to the same scientific computation domain
(optimization, linear algebra, differential equations,~etc). Another
property that is usually shared by all the problems of a collection is
the extend to which they are protected against dissemination; this is
useful for determining the public to which the collection can be
distributed.

A {\em subcollection\/}\index{subcollection} is a subset of problems
belonging to the same collection. Its {\em
name\/}\index{subcollection!name of a --} must also be match by
\texttt{\char`\\w+}. A collection may contain several subcollections,
and a problem may belong to more than one subcollection. The reason why
the notion of subcollection is introduced is clear: some solvers can
sometimes only solve a part of the problems of a given collection and
it is useful to be able to designate them. For example, in
optimization, there is some interest in distinguishing the
subcollections of unconstrained problems, of linear problems, of
quadratic problems, of bound constrained problems, etc, since there are
solvers that are dedicated to these classes of problems.

\subsection{Running solvers with \texttt{runopt}}
\label{s:runopt-basic}

\index{command (Libopt --)!runopt@\texttt{runopt}|(}

Libopt can only deal with solvers that are registered in its
environment. The procedure to do this is described in
section~\ref{s:solver}. Below, we denote by
\begin{quote}%
\index{solver!solv generic name@\SOLV\ (generic name)}%
\SOLV
\end{quote}
the generic name of such a solver. Similarly, Libopt can only consider
collections of problems that are installed in its hierarchy. This
simply means that some directory (or a symbolic link to it) has to
contain the problems of the collection (these can be in an arbitrary
format) and some files and scripts have to tell how to use the
collection (this is why the format can be arbitrary). This subject is
discussed in section~\ref{s:adding-collection}. Below, we denote by
\begin{quote}%
\index{collection!coll generic name@\COLL\ (generic name)}%
\COLL, \SUBC, and \PROB
\index{collection!coll generic name@\COLL\ (generic name)}
\index{subcollection!subc generic name@\SUBC\ (generic name)}
\index{problem!prob generic name@\PROB\ (generic name)}
\end{quote}
the generic names of a collection, subcollection, and problem,
respectively.

The simplest way of running the solver \SOLV\ on the problem \PROB\ of
the collection \COLL\ in the Libopt environment is by entering
\begin{quote}\label{first-runopt}%
\texttt{%
\% echo "\SOLV\ \COLL\ \PROB" | runopt}%
\index{command (Unix/Linux --)!echo@\texttt{echo}}%
\end{quote}%
In other words, the standard input of the script \texttt{runopt} is fed
with the {\em command string\/} ``\SOLV\ \COLL\ \PROB''. We quote a few
interests in proceeding like this.
\begin{list}{{\small$\bullet$}}{\topsep=1.5ex\parsep=0.5ex\itemsep=0.5ex
 \settowidth{\labelwidth}{{\small$\bullet$}}
 \labelsep=0.5em
 \leftmargin=\bulindent
}
\item
First, the form of the command for running any solver on any collection
is invariant: it does not depend on the solver, the collection, or the
problem. In our experience, this property saves much memorization
effort. In particular, it gives to a collection, which is not used
often and is coded in a manner that is difficult to remember, more
chance to be tested, even after it has been abandoned for several
years.

\item
Second, it defines a standard, to which solver developers can
contribute by providing the interfaces between their solver and various
collections.

\item
Also, the possibility to consider a large diversity of collections
should allow the environment to accept problems coming from various
sources.

\end{list}

A slightly more powerful use of the \texttt{runopt} command is
\begin{quote}
\texttt{%
\% echo "\SOLV\ \COLL.\SUBC" | runopt}%
\index{command (Unix/Linux --)!echo@\texttt{echo}}
\end{quote}
where \SUBC\ is a subcollection of the collection \COLL. By this
command the solver \SOLV\ is run on all the problems of the
subcollection \SUBC. If the suffix ``\texttt{.\SUBC}'' is not present
in the command string above, the \texttt{all} or \texttt{default}
subcollection is assumed, depending on the presence or absence of
problems in the command string. Subcollections are described by lists
of problems (see section~\ref{s:lists}), which are searched in various
directories by \texttt{runopt} (see section~\ref{s:runopt}).

\index{command (Libopt --)!runopt@\texttt{runopt}|)}

\subsection{Gathering the results with \texttt{addopt}}
\label{s:addopt-basic}

\index{command (Libopt --)!addopt@\texttt{addopt}|(}

By the \texttt{runopt} command a solver-problem pair writes its output
on its usual files, which probably include the Unix/Linux standard
output. In order to compare solvers, there is no reason to save all
these files, which mainly interest the developer of the problem code.
In fact, the standard Libopt scripts normally remove these files after
having run a solver on a problem (this behavior can be prevented by
setting the option \texttt{-k} of the \texttt{runopt} command, see the
introduction of section~\ref{s:depths} and
section~\ref{s:runopt-script}). This is because, Libopt is designed to
{\em compare\/} the results of various solvers, not to {\em analyze\/}
them. For this reason, Libopt is interested in the value of various
counters that reflect the performance of a solver on a particular
problem. In optimization, these counters are often the number of
function or derivative evaluations, the CPU time, the precision of the
obtained solution, and many others.

In order to be able to make comparisons, Libopt imposes that the
results relevant to a comparison be condensed on the standard output
in a string of the form
\begin{quote}
\label{libopt-line}
\texttt{%
libopt\%\SOLV\%\COLL\%\PROB\%\textit{sequence-of-token-number-pairs}}
\end{quote}
This one is called the {\em Libopt line}\index{Libopt line|ibf}. It is
formed of a sequence of fields separated by the
character~`\texttt{\%}'.
\begin{list}{{\small$\bullet$}}{\topsep=1.5ex\parsep=0.5ex\itemsep=0.5ex
 \settowidth{\labelwidth}{{\small$\bullet$}}
 \labelsep=0.5em
 \leftmargin=\bulindent
}
\item
The first field is the string ``\texttt{libopt}''. It is present to
make it easier to locate the Libopt line in the standard output (using
the command \texttt{grep}\index{command (Unix/Linux
--)!grep@\texttt{grep}} for example). Therefore this string must appear
only once in the standard output and the Libopt line cannot be split on
several lines on the standard output.

\item
The next three fields give in order the name of the solver (\SOLV), the
name of the collection (\COLL), and the name of the problem (\PROB),
whose relevant results are given in the following fields.
\end{list}

\noindent
These first four fields are positional, i.e., their order is imposed.
This is not the case for the following ones.

\begin{list}{{\small$\bullet$}}{\topsep=1.5ex\parsep=0.5ex\itemsep=0.5ex
 \settowidth{\labelwidth}{{\small$\bullet$}}
 \labelsep=0.5em
 \leftmargin=\bulindent
}
\item
The \texttt{\textit{sequence-of-token-number-pairs}} is a string formed
of a sequence of token-number pairs, again separated by the
character~`\texttt{\%}'. A {\em token-number pair\/}\index{token-number
pair} is a string of the form
\begin{quote}
\texttt{%
\textit{token}=\textit{number}}
\end{quote}
The character `\texttt{=}' must separate the {\em token\/}\index{token}
(a string matched by \texttt{\char`\\w+}) from the {\em number\/}
(a string representing a real number). There must be at least two
token-number pairs, one of which is used to compare the results (it
must be a {\em performance\/}\index{token!performance} token-number
pair actually, see below) and another one must have the form
\begin{quote}
\texttt{%
info=\textit{number}}
\end{quote}
where the string ``\texttt{info}''\index{token!info@\texttt{info}}
is imposed and a \texttt{\textit{number}} equal to \texttt{0} means
that the solver \SOLV\ has successfully solved the problem \PROB.

Note that, since the Libopt line is written by some program provided by
the developer of a solver, it is that program that decides whether the
solver has successfully solved a given problem; Libopt has no means to
take such a decision.
\end{list}
The libopt line can be written with some flexibility: blanks (matched
by \texttt{\char`\\u+}) surrounding the various fields and elements of
fields are ignored, and a comment\index{Libopt line!comment in the --}
can be introduced (it starts with the sharp character~`\texttt{\#}' and
goes up to the end of the line).

An example of Libopt line in optimization could be
\begin{quote}
\label{libopt-line-in-optimization}
\texttt{%
libopt\%m1qn3\%modulopt\%u1mt1\%n=1875\%nfc=143\%nga=143\%info=0}
\end{quote}
to mean that the solver
\texttt{m1qn3}\index{solver!m1qn3@\texttt{m1qn3}} has been run on the
problem \texttt{u1mt1} of the collection \texttt{modulopt}, that this
problem has 1875 variables (\texttt{n=1875}) and that a solution has
been found (\texttt{info=0}) using 143 function and gradient
evaluations (\texttt{nfc=143} and \texttt{nga=143}).

Now, the command
\begin{quote}
\index{command (Libopt --)!runopt@\texttt{runopt}}
\texttt{%
\% echo "\SOLV\ \COLL.\SUBC" | runopt | grep libopt
\char`\\}\index{command (Unix/Linux
--)!grep@\texttt{grep}}\index{command (Unix/Linux
--)!echo@\texttt{echo}}\\
\hphantom{\texttt{\% }}\texttt{%
> \SOLV\_\COLL\_\SUBC.lbt}
\end{quote}
gathers in the file \texttt{\SOLV\_\COLL\_\SUBC.lbt} a sequence of
Libopt lines describing the behavior of the solver \SOLV\ on the
problems of the subcollection~\SUBC\ of the collection~\COLL. It is
interesting to save this file preciously, since the previous command
may have required much time to run, since it is not a large file, and
since its ascii encoding makes it very stable with respect to the
possible evolutions of the languages (Libopt, Perl, or Unix/Linux).

There may be many files of results like \texttt{\SOLV\_}\abr
\texttt{\COLL\_}\abr \texttt{\SUBC.}\abr \texttt{lbt} and there is some
interest in gathering all the results they contain in a single file.
This gathering operation is also a good opportunity to verify that the
Libopt lines in the result files are consistent and that there is at
most one result for each (solver, collection, problem) triple. This is
exactly what the command \texttt{addopt} does. If one enters
\begin{quote}
\index{command (Libopt --)!addopt@\texttt{addopt}}
\texttt{%
\% addopt \SOLV\_\COLL\_\SUBC.lbt}
\end{quote}
the Libopt lines in the result file are decoded, checked (see below),
and stored in a database, whose default name is
\texttt{dtbopt}\index{file!dtbopt default database name@\texttt{dtbopt}
(default database name)} in the working directory. This database is no
longer an ascii file, but a binary file or a pair of binary files,
depending on the operating system. Possible file names are
\texttt{dtbopt.}\abr \texttt{db} or the pair \texttt{dtbopt.}\abr
\texttt{dir}/\abr \texttt{dtbopt.}\abr \texttt{pag}. This variety of
storages makes the database not very portable, which is also a reason
to save the ascii files that have been used to generate it. The
database stores a collection of key-value pairs: the key is the first
part of the Libopt line, more exactly the string ``\SOLV\abr
\texttt{\%}\COLL\abr \texttt{\%}\PROB'', without useless blanks (there
is no reason to store the invariant string ``\texttt{libopt\%}''); the
value is the second part of the Libopt line, namely what we have
denoted the \texttt{\textit{sequence-}}\abr \texttt{\textit{of-}}\abr
\texttt{\textit{token-}}\abr \texttt{\textit{number-}}\abr
\texttt{\textit{pairs}} above. By entering the \texttt{addopt} command
for all the relevant result files, one can obtain a database containing
all the results of interest, without duplicated or contradictory data.
The database can be managed, using the options of the \texttt{addopt}
command: \texttt{-r} for replacing entries and \texttt{-d} for deleting
results. A full description of the \texttt{addopt} command line is
given in section~\ref{s:addopt} and in its manual page.

The amount of verifications done  on the libopt lines by
\texttt{addopt} depends on the presence and contents of the
\texttt{\char`~/.liboptrc}\index{file!.liboptrc startup
file@\texttt{.liboptrc} (startup file)} startup file (see
section~\ref{s:liboptrc} for a complete description of this file). If
there is no such file, \texttt{addopt} just verifies the conventions
mentioned in the beginning of this section. It cannot do more. In
particular, it makes no assumption on what are the tokens, on the
quantity of token-number pairs, and (clearly) on their order. These
pieces of information are actually very domain dependent and certainly
not identical in optimization, in linear algebra, or in the
differential equation domain. However, to avoid typos, you can list the
valid tokens in the startup file \texttt{\char`~/.tcshrc}. If this file
is present, \texttt{addopt} will read it, and if it contains the
directive \texttt{tokens}, \texttt{addopt} will verify that the tokens
in the Libopt line are among those listed by this directive.

\index{command (Libopt --)!addopt@\texttt{addopt}|)}

\subsection{Comparing the results with \texttt{perfopt}}
\label{s:comparing-results}

\index{command (Libopt --)!perfopt@\texttt{perfopt}|(}

The \texttt{perfopt} command has been designed to make comparisons
between solvers on a selected set of problems. The comparison is based
on the results stored in a database, one generated by the
\texttt{addopt} command. A single criterion can be used for making this
comparison, and it must be one of the tokens present in the Libopt
lines\index{Libopt line} summarizing the results to compare. This
comparison made by \texttt{perfopt} produces files that can be
subsequently dealt with Matlab or Gnuplot. These files describe the
{\em performance profiles\/}\index{performance!profile} \`a la Dolan
and Mor\'e~\cite{dolan-more-2002} of the solvers.

In the Libopt line, one finds descriptive token-number pairs (or
descriptive tokens) and performance ones. The semantics behind this
distinction is rather intuitive: a {\em performance
token\/}\index{token!performance|ibf} is a token that can be used to
compare solvers, while a {\em descriptive
token}\index{token!descriptive} is a non-performance token. A
performance token-number pair must have the following properties:
\begin{list}{{\small$\bullet$}}{\topsep=1.5ex\parsep=0.0ex\itemsep=0.0ex
 \settowidth{\labelwidth}{{\small$\bullet$}}
 \labelsep=0.5em
 \leftmargin=\bulindent
}
\item
the token-number pair must, obviously, depend on the solver (not on the
problem),
\item
the number of the token-number pair must be positive ($>0$),
\item
the number of the token-number pair is {\em better\/} (i.e., indicates
a better performance) when it is {\em smaller\/}.
\end{list}
In the example given on page \pageref{libopt-line-in-optimization},
\texttt{n=1875} must be considered as a descriptive token-number pair
(the dimension of the problem is not a property of the solver), while
\texttt{nga=143} is normally a performance one, since an optimization
solver can be considered to be better when it requires less derivative
evaluations. Libopt cannot verify that these rules have been respected
(except for the positivity of the number), but the performance profiles
that \texttt{perfopt} generates assume that they hold; they would have
no meaning otherwise. On the other hand, if you want Libopt to make
some verifications on the correct use of performance tokens, you can
list them in the startup file \texttt{\char`~/.tcshrc} (see
section~\ref{s:liboptrc} for more precision on this file).

Note, however, that descriptive tokens may also be interesting for
selecting the problems on which a comparison is made. For example, one
can be interested in making a comparison of solvers on problems with
more (or less) than $1000$ variables, and/or those with (or without)
inequality constraints, etc. To see how to prescribe this, read what is
said about the ``problem directive'' in the description of the
\texttt{perfopt.}\abr \texttt{spc}\index{file!perfopt
spc@\texttt{perfopt.}\abr \texttt{spc} (specification file)}
specification file in section~\ref{s:perfopt} or in the manual page of
\texttt{perfopt}.

\begin{figure}[htbp]
\centering
\psfrag{goya}{\footnotesize\texttt{{bosch}}}
\psfrag{klee}{\footnotesize\texttt{{klee}}}
\psfrag{durerxxx}{\footnotesize\texttt{{durer}}}
\includegraphics[scale=0.5]{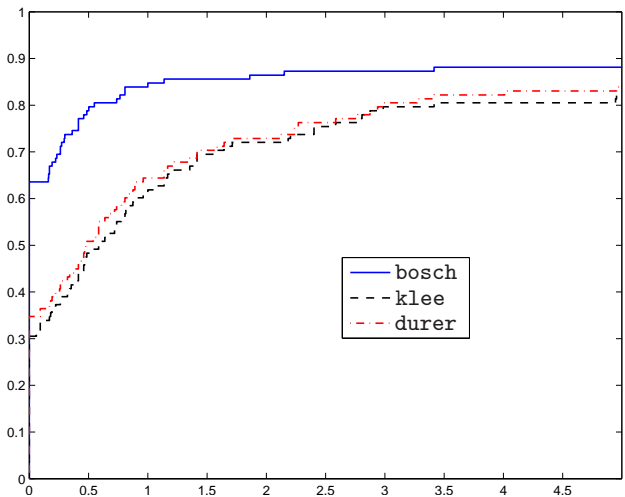}
\caption{\label{f:pp}Typical performance profiles for three solvers}
\end{figure}

The easiest way of using the script \texttt{perfopt} is by entering
\begin{quote}
\index{command (Libopt --)!addopt@\texttt{addopt}}
\texttt{\% perfopt}
\end{quote}
Then, the script scrutinizes the database (its default name is
\texttt{dtbopt} in the working directory) and reads the specification
file \texttt{perfopt.}\abr \texttt{spc}\index{file!perfopt
spc@\texttt{perfopt.}\abr \texttt{spc} (specification file)} in the
working directory to know what it has to do. This file is described in
detail in section~\ref{s:perfopt}. It can contain many {\em
directives}\index{directive!perfopt@\texttt{perfopt}}. Three of the
most important ones are those that specify the solvers that have to be
compared, the problems on which they have to be compared, and the
performance token chosen for the comparison. For example
\begin{quote}
\texttt{%
solver \texttt{bosch} \texttt{klee} \texttt{durer}}\\
\texttt{%
collection painting.selfportraiture}\\
\texttt{%
performance time}
\end{quote}
is understood by \texttt{perfopt} as a requirement to compare the
execution time (if this is the meaning of the token \texttt{time} in
the third line) for the three solvers \texttt{bosch}, \texttt{klee},
and \texttt{durer} (first line), on the problems from the subcollection
\texttt{selfportraiture} of the collection \texttt{painting} (second
line). Then \texttt{perfopt} generates files with the default names
\texttt{perf.}\abr \texttt{gnu} and \texttt{perf.m}, which subsequently
processed by Gnuplot or Matlab can give a graph like the one in
figure~\ref{f:pp}
(a few additional tunings in the specification file are necessary to
get precisely this graph).

\index{performance!profile|(ibf}

{\em Performance profiles\/} have been introduced by Dolan and
Mor\'e~\cite{dolan-more-2002}. To be comprehensive, we feel it
necessary to give a few words on the meaning of these curves. More can
be found in the original paper. The main objective is to replace tables
of numbers by curves (one per solver), which, with some reading-keys,
provide a rapid understanding of the {\em relative\/} performance of
various solvers on a given set of problems. More specifically, these
curves are used to compare the efficiency of a set ${\cal S}$ of
solvers on a set ${\cal P}$ of test problems. Let
$$
\tau_{p,s} := \mbox{performance of the solver $s$ on the problem $p$,}
$$
where the {\em performance\/}\index{performance} refers to the number
of a token-number performance pair. The {\em relative
performance\/}\index{performance!relative} of a solver~$s$ (with
respect to the other solvers in ${\cal S}$) on a problem~$p$ is the
ratio
$$
\rho_{p,s}=\frac{\tau_{p,s}}{\min\{\tau_{p,s'}:s'\in{\cal S}\}}.
$$
Of course $\rho_{p,s}\geq1$. On the other hand, it is assumed that
$\rho_{p,s} \leq \bar{\rho}$ for all problems~$p$ and solvers~$s$, which
can be ensured only by setting $\rho_{p,s}$ to the large number
$\bar{\rho}$ if the solver~$s$ cannot solve the problem~$p$
(\texttt{info} is nonzero in the Libopt line\index{Libopt line}).
Actually, we shall consider that~$s$ fails to solve~$p$ if and only if
$\rho_{p,s}=\bar{\rho}$. The {\em performance profile\/} of the
solver~$s$ (relative to the other solvers) is then the function
$$
t\in[1,\bar{\rho}]\mapsto\wp_s(t):=
\frac{|\{p\in{\cal P}:\rho_{p,s}\leq t\}|}{|{\cal P}|}\in[0,1],
$$
where $|\cdot|$ is used to denote the number of elements of a set (its
cardinality).

Only three facts need to be kept in mind to have a good interpretation
of these upper-semi-continuous piecewise-constant nondecreasing
functions:
\begin{list}{{\small$\bullet$}}{\topsep=1.5ex\parsep=0.5ex\itemsep=0.5ex
 \settowidth{\labelwidth}{{\small$\bullet$}}
 \labelsep=0.5em
 \leftmargin=\bulindent
}
\item
$\wp_s(1)$ gives the fraction of problems on which the solver~$s$ is
the best; note that two solvers may have an even score and that all the
solvers may fail to solve a given problem, so that it is not guaranteed
to have $\sum_{s\in{\cal S}}\wp_s(1)=1$;

\item
by definition of $\bar{\rho}$, $\wp_s(\bar{\rho})=1$; on the other
hand, for small $\varepsilon>0$, $\wp_s(\bar{\rho}-\varepsilon)$ gives
the fraction of problems that the solver~$s$ can solve; this value is
independent of the performance token chosen for the comparison;

\item
the value $\wp_s(t)$ may be given an interpretation by inverting the
function $t\mapsto\wp_s(t)$: for the fraction $\wp_s(t)$ of problems in
${\cal P}$, the performance of the solver~$s$ is never worse than $t$
times that of the best solver (this one usually depends on the
considered problem); in this respect the argument at which $\wp_s$
reaches its ``almost maximal'' value $\wp_s(\bar{\rho}-\varepsilon)$ is
meaningful.
\end{list}
With performance profiles, the relative efficiency of each solver
appears at a glance: the higher is the graph of $\wp_s$, the better is
the solver~$s$.

\index{performance!profile|)ibf}

\index{command (Libopt --)!perfopt@\texttt{perfopt}|)}

\section{The Libopt package}
\label{s:install}

This section contains a brief explanation on how to unpack and install
the Libopt package, and a rather detailed discussion of its
directory/file structure.

\subsection{Retrieving the environment}

Libopt has been designed to be used by more than one person at the same
time, so that it is recommended to put the Libopt hierarchy in a
location that is visible to all the potential users. The Libopt
hierarchy can be retrieved by one of the following two manners.
\begin{list}{{\small$\bullet$}}{\topsep=1.5ex\parsep=0.5ex\itemsep=0.5ex
 \settowidth{\labelwidth}{{\small$\bullet$}}
 \labelsep=0.5em
 \leftmargin=\bulindent
}
\item
The first possibility is to get the package from the Inria GForge
\texttt{http:}\abr \texttt{//gforge.}\abr \texttt{inria.}\abr
\texttt{fr}, using the \texttt{svn} command
\begin{quote}
\index{command (Unix/Linux --)!svn@\texttt{svn}}
\texttt{%
\% svn checkout svn+ssh://\textit{name}@scm.gforge.inria.fr/svn/libopt}
\end{quote}
where \texttt{\textit{name}} stands for a registered name. This method
requires to be registered to the GForge and to have set up the
\texttt{ssh} communication. It is therefore more appropriate for the
developers of (and the contributers to) Libopt.

\item
The other possibility is to retrieve the tar-gzip file
\begin{quote}
\texttt{%
libopt.tar.gz}
\end{quote}
and to unpack it, using successively the commands
\texttt{gunzip}\index{command (Unix/Linux --)!gunzip@\texttt{gunzip}}
and
\texttt{tar}\index{command (Unix/Linux --)!tar@\texttt{tar}}.
\end{list}

%
%
%
%
%
%

\subsection{Structure of the Libopt hierarchy}

The retrieving procedure above creates a tree of subdirectories, whose
root name is \texttt{libopt}. This top-level directory contains the
directories \texttt{bin} (Perl scripts of the Libopt commands),
\texttt{collections} (problem collections), \texttt{doc} (some
documentation), \texttt{man} (manual pages), \texttt{platforms}
(platform descriptions, see section~\ref{s:platforms}), and
\texttt{solvers} (solver descriptions and interfaces); see
figure~\ref{f:libopt-hierarchy}.
\begin{figure}[htbp]
\begin{center}
\newlength{\verbatimwidth}
\settowidth{\verbatimwidth}{\texttt{xxxxxxxxxxxxxxxxxxxxxxxxxxxxxxxxxxxxxxxxxxxxxxxxxxxxxxxxxxxxxxxx}}
\raisebox{+0.0ex}{\rule{\verbatimwidth}{0.5pt}}
\begin{minipage}{\verbatimwidth}
\begin{verbatim}
Level-0      Level-1      Level-2      Description
\end{verbatim}
\end{minipage}
\raisebox{+1.5ex}{\rule{\verbatimwidth}{0.5pt}}
\begin{minipage}{\verbatimwidth}
\index{solver!m1qn3@\texttt{m1qn3}}
\index{solver!sqppro@\texttt{sqppro}}
\begin{verbatim}
bin .................................. Libopt scripts
collections
        +--- cuter ................... CUTEr collection
        +--- modulopt ................ Modulopt collection
doc .................................. Documentation
man     +--- man1 .................... Manual pages
platforms ............................ Platform descriptions
solvers +--- m1qn3   +--- bin ........ M1qn3 binary directory
        |            +--- cuter ...... Interface M1qn3/CUTEr
        |            +--- modulopt ... Interface M1qn3/Modulopt
        +--- sqppro  +--- bin ........ Sqppro binary directory
                     +--- cuter ...... Interface Sqppro/CUTEr
                     +--- modulopt ... Interface Sqppro/Modulopt
\end{verbatim}
\end{minipage}
\raisebox{-1ex}{\rule{\verbatimwidth}{0.5pt}}
\end{center}
\caption{\label{f:libopt-hierarchy}Part ot the Libopt hierarchy in the
standard distribution}
\end{figure}

The \texttt{collections} directory gathers a set of subdirectories,
each of them corresponding to an installed collection of problems. In
the standard distribution, one finds the two collections
CUTEr\index{collection!CUTEr}~\cite{gould-orban-toint-2003} in
\texttt{cuter} and
Modulopt\index{collection!Modulopt}~\cite{lemarechal-1980,
gilbert-2007b} in \texttt{modulopt}. More generically, the directory
\texttt{collections/}\abr \texttt{\COLL} contains the description of
the collection \COLL. What this actually means is reflected in the
scripts that use this collection, so that these directories can be
organized with a great freedom, as far as Libopt is concerned. More is
said about this directory in section~\ref{s:adding-collection}.

It is natural to put in these collection directories (the
subdirectories of \texttt{col}\abr \texttt{lections}) some {\em lists
of problems}\index{list!of problems}. These are files whose name must
have the suffix ``\texttt{.lst}''\index{suffix!.lst@\texttt{.lst}} (see
also section~\ref{s:lists}). Two of these lists are mandatory:
\begin{list}{{\small$\bullet$}}{\topsep=1.5ex\parsep=0.0ex\itemsep=0.0ex
 \settowidth{\labelwidth}{{\small$\bullet$}}
 \labelsep=0.5em
 \leftmargin=\bulindent
}
\item
\texttt{all.lst}\index{list!all.lst@\texttt{all.lst}|ibf} is the
list of all the problems of the collection; \item
\texttt{default.lst}\index{list!default.lst@\texttt{default.lst}|ibf}
is a list of a subset of the problems of the collection; this list is
chosen by some scripts when no list is specified as one of their
arguments (see section~\ref{s:runopt-script}); it is sometimes a
symbolic link to the list \texttt{all.}\abr \texttt{lst}.
\end{list}
It is also natural to have in \texttt{collections/}\abr \texttt{\COLL}
a subdirectory, usually called \texttt{probs}, which contains all the
problems of the collection. Of course \texttt{probs} can be a symbolic
link to the actual problem directory. For example, in our Libopt
hierarchy, \texttt{collections/}\abr \texttt{cuter/}\abr \texttt{probs}
is a symbolic link to a directory containing the gzipped SIF encoding
of the CUTEr problems. On the other hand, in the standard distribution,
\texttt{collections/}\abr \texttt{modulopt/}\abr \texttt{probs}
contains a subdirectory for each of the problems of the Modulopt
collection, which describes this problem by programs, data, and a
makefile.

%
%

The directory \texttt{collections} provides no information on the
solvers. This information is given by the directory \texttt{solvers},
which has a subdirectory for each solver known to the Libopt
environment: \texttt{m1qn3}\index{solver!m1qn3@\texttt{m1qn3}} and
\texttt{sqppro}\index{solver!sqppro@\texttt{sqppro}} in the standard
distribution. \texttt{M1qn3} is an unconstrained optimization solver,
using the $\ell$-BFGS formula~\cite{gilbert-lemarechal-1989}, and
\texttt{sqppro} is a nonlinear optimization solver, using an SQP
approach~\cite{delbos-gilbert-2005,
delbos-gilbert-glowinski-sinoquet-2006}. The directory \texttt{solvers}
is actually very rich, since, in some sense, it reflects the structure
of the Cartesian product 
\begin{equation}
\label{cartesian-product}
\{\texttt{solv\_1}, \ldots, \texttt{solv\_m}\} \times 
\{\texttt{coll\_1}, \ldots, \texttt{coll\_n}\}
\end{equation}
corresponding to the interfaces between solvers and collections.
Indeed, for each pair (\SOLV,\COLL), where \SOLV\ is some solver
\texttt{solv\_i} and \COLL\ is some collection \texttt{coll\_j}, the
directory \texttt{solvers/}\abr \texttt{\SOLV/}\abr \texttt{\COLL}
contains information describing how to run the solver \SOLV\ on the
problems of the collection \COLL. This directory includes the following
mandatory files/scripts (see also figure~\ref{f:cartesian-product}):
\begin{figure*}[htbp]
\begin{center}
 \input{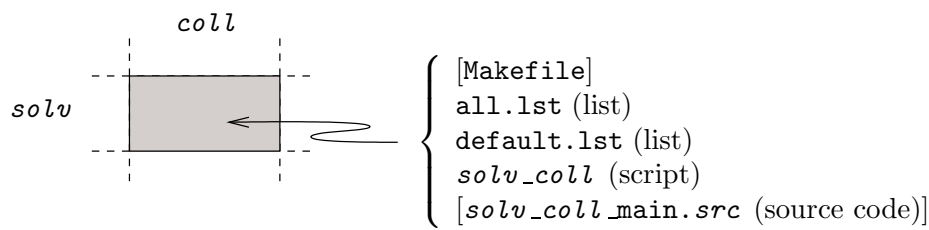}
\end{center}
\caption{\label{f:cartesian-product}%
A single cell of the $\{$solvers$\}\times\{$collections$\}$ Cartesian
product}
\end{figure*}
\begin{list}{{\small$\bullet$}}{\topsep=1.5ex\parsep=0.5ex\itemsep=0.5ex
 \settowidth{\labelwidth}{{\small$\bullet$}}
 \labelsep=0.5em
 \leftmargin=\bulindent
}
\item
\texttt{all.lst}\index{list!all.lst@\texttt{all.lst}|ibf} is the
list of all the problems of the collection \COLL\ that the solver
\SOLV\ is structurally able to solve (note that this meaning is quite
different from the one of the file \texttt{collections/}\abr
\texttt{\COLL/}\abr \texttt{all.lst} described above); for instance,
the file \texttt{solvers/}\abr \texttt{m1qn3/}\abr \texttt{cuter/}\abr
\texttt{all.lst} is a symbolic link to the list of unconstrained
problems of the CUTEr collection;

\item
\texttt{default.lst}\index{list!default.lst@\texttt{default.lst}|ibf}
lists a subset of the problems in \texttt{all.lst}; this list is used
by some scripts when no list is specified as one of its arguments (see
section~\ref{s:runopt-script});

\item
\texttt{\SOLV\_\COLL}\index{command (Libopt --)!solv
coll@\texttt{\SOLV\_}\abr \texttt{\COLL}} is a script that tells how to
run the solver \SOLV\ on a problem of the collection \COLL; it
essentially takes care of the operating system commands that are
required to launch \SOLV\ (see section~\ref{s:solv-coll-script}); note
that the name of the script depends on the directory where it is
placed.
\end{list}
The directory may also contain other files, like the source code of a
main program \texttt{\SOLV\_}\abr \texttt{\COLL\_}\abr
\texttt{main.}\abr \texttt{\textit{src}} able to run the solver \SOLV\
on a single problem of the collection \COLL, which takes care of the
instructions that cannot be written in a Unix/Linux script, and a
makefile \texttt{Makefile} (see section~\ref{s:main}).


\subsection{Installation instructions}

\index{installation instructions|(}

Once the Libopt has been unpacked, a few things need to be done before
being able to use the environment. We suggest the following five steps.

\begin{list}{}{\topsep=0.5ex\partopsep=0.5ex\parsep=0.5ex\itemsep=0.5ex
 \settowidth{\labelwidth}{9.}
 \labelsep=0.5em
 \leftmargin=\labelwidth
 \addtolength{\leftmargin}{\labelsep}
}
\item[1.]
First, add the following environment variable definitions to the
shell startup file \texttt{\char`~/.tcshrc} (the \texttt{tcsh} shell is
assumed; adapt the format and use another startup file if the shell is
different; echoing your \texttt{\$SHELL}\index{environment
variable!SHELL@\texttt{SHELL}} environment variable might be useful
here):
\begin{quote}
\texttt{%
setenv LIBOPT\_DIR~~~~\textit{dir}}\\
\texttt{%
setenv LIBOPT\_PLAT~~~\textit{plat}}\\
\texttt{%
setenv PATH~~~~~~~~~~\$\char`{PATH\char`}:\$\char`{LIBOPT\_DIR\char`}/bin}
\end{quote}
The variable \texttt{\$LIBOPT\_}\abr \texttt{DIR}\index{environment
variable!LIBOPT DIR@\texttt{LIBOPT\_}\abr \texttt{DIR}} sets the root
directory of the Libopt hierarchy; hence change
``\textit{\texttt{dir}}'' into the location of this root directory. The
variable \texttt{LIBOPT\_}\abr \texttt{PLAT}\index{environment
variable!LIBOPT PLAT@\texttt{LIBOPT\_}\abr \texttt{PLAT}} provides the
{\em platform\/}\index{platform} on which you work, see
section~\ref{s:platforms} for valid value for the string
``\textit{\texttt{plat}}''. The third setting adds to \texttt{\$PATH}
the directory of the Libopt commands. In some Linux system, this
setting automatically adds \texttt{\${LIBOPT\_}\abr \texttt{DIR}/man}
to the possible search paths for the manual pages. If this is not the
case with your system, you might have to set
\begin{quote}
\begin{verbatim}
setenv MANPATH       ${MANPATH}:${LIBOPT_DIR}/man
\end{verbatim}
\end{quote}
With these settings you should have an appropriate answer to the
following commands:
\begin{quote}
\begin{verbatim}
% runopt -h
% man runopt
\end{verbatim}
\end{quote}

\item[2.]
You can now launch the installation procedure (from any directory,
since this script is in \texttt{\${LIBOPT\_DIR}/}\abr \texttt{bin}):
\begin{quote}
\index{command (Libopt --)!install libopt@\texttt{install\_libopt}}
\begin{verbatim}
% install_libopt
\end{verbatim}
\end{quote}
This procedure defines the files \texttt{collections/}\abr
\texttt{collections.}\abr \texttt{lst} (the list of
collections\index{list!of collections} existing in the standard
distribution), \texttt{solvers/}\abr \texttt{solvers.}\abr \texttt{lst}
(the list of solvers\index{list!of solvers} existing in the standard
distribution), and for each solver \SOLV\ in the previous list, it
defines \texttt{solvers/}\abr \texttt{\SOLV/}\abr
\texttt{collections.}\abr \texttt{lst} (the list of
collections\index{list!of collections} that the solver \SOLV\ has been
prepared to consider). It also verifies the consistency of the
Libopt hierarchy, by checking whether expected files are present.

\item[3.]
This third step is optional but recommended. It consists in introducing
the startup file \texttt{\char`~/.liboptrc}\index{file!.liboptrc
startup file@\texttt{.liboptrc} (startup file)} (hence in your home
directory). This file provides additional information to some Libopt
commands, for example for helping \texttt{addopt} to detect typos in
Libopt lines. See section~\ref{s:liboptrc} for a detailed description
of the directives that can be put in this startup file.

In optimization, one can start by copying a file given in the
\texttt{doc} directory
\begin{quote}
\begin{verbatim}
% cp $LIBOPT_DIR/doc/liboptrc_optim ~/.liboptrc
\end{verbatim}
\end{quote}
and modify it afterwards if the tokens used in the file does not suit
to ones needs.



\item[4.]
The fourth step deals with the installation of a collection. The
Modulopt collection is already installed in the Libopt hierarchy, so
that nothing need to be done for being able to use it. Very little has
to be done for the installation of the CUTEr collection: follow the
procedure detailed in section~\ref{s:hooking-cuter-collection}.

\item[5.]
This fifth step is the most complex. It deals with the installation of
a solver in the Libopt environment. Indeed, Libopt does not provide
solvers.

If you are lucky, you might have one of the solvers of the standard
distribution, those now listed in the file \texttt{solvers/}\abr
\texttt{solvers.}\abr \texttt{lst}\index{list!of solvers}. In this
case, the installation is usually rapid; it is described in the file
\texttt{solvers/}\abr \texttt{\SOLV/}\abr \texttt{README\_}\abr
\texttt{install} if \SOLV\ is the considered solver. Otherwise, follow
the steps given in section~\ref{s:solver}.
\end{list}

You should now be able to run a solver (say \SOLV) on a problem (say
\PROB) of some collection (say \COLL), by typing
\begin{quote}
\texttt{%
\% echo "\SOLV\ \COLL\ \PROB" | runopt}%
\index{command (Unix/Linux --)!echo@\texttt{echo}}
\end{quote}
The next section gives more details on the possibilities of the Libopt
environment and explains how the commands are dealt with.

\index{installation instructions|)}

\section{Libopt in depth}
\label{s:depths}

Section~\ref{s:tour} has presented the tools the most frequently used
of the Libopt environment. Sometimes, however, it is necessary to do
other operations like introducing a new collection of problems or a new
solver in the environment. A deeper understanding of the Libopt
mechanisms is required for realizing these operations safely. The goal
of this section is to provide a comprehensive description of the
principles governing the software, as well as the available commands.

The Libopt commands are located in the \texttt{bin} directory of the
hierarchy. They are written in Perl. Many of these commands accept
several of the following options, whose meaning will not be repeated:
\begin{list}{}{\topsep=0.5ex\partopsep=0.5ex\parsep=0.0ex\itemsep=0.0ex
 \settowidth{\labelwidth}{\texttt{-h}}
 \labelsep=1.0em
 \leftmargin=\labelwidth
 \addtolength{\leftmargin}{\labelsep}
 \addtolength{\leftmargin}{1.0em}
}
\item[\texttt{-h}\index{h@\texttt{-h} (command option)}]
help mode; a short help message describing the command usage is
printed;

\item[\texttt{-k}\index{k@\texttt{-k} (command option)}]
keep mode; the files generated by the command and its children are not
removed on exit;

\item[\texttt{-t}\index{t@\texttt{-t} (command option)}]
test mode; same as \texttt{-v}, but the commands are not executed;

\item[\texttt{-v}\index{v@\texttt{-v} (command option)}]
verbose mode; the commands are executed and are also printed on the
standard output.
\end{list}
When a command launches other scripts and some of the options
\texttt{-k}, \texttt{-t}, and/or \texttt{-v} are present, these options
are transmitted to these scripts.

\subsection{The \texttt{\char`~/.liboptrc} startup file}
\label{s:liboptrc}

\index{file!.liboptrc startup file@\texttt{.liboptrc} (startup file)|(}

The file \texttt{\char`~/.liboptrc} is used to provide some additional
information to Libopt, in connection with the activity of the user.
Here are the {\em
directives\/}\index{directive!liboptrc@\texttt{.liboptrc}} that are
meaningful:
\begin{quote}
\texttt{%
tokens = \textit{list-of-tokens}}\\
\texttt{%
performance\_tokens = \textit{list-of-performance-tokens}}
\\
\texttt{%
data\_base = \textit{DBFile}}
\end{quote}

The first two directives are related to the Libopt line (see
page~\pageref{libopt-line}). The \texttt{\textit{list-}}\abr
\texttt{\textit{of-}}\abr \texttt{\textit{tokens}} is a blank-separated
list of the tokens that are considered to be valid in the Libopt line.
If this directive is present, \texttt{addopt} uses the tokens of this
list to determine whether the tokens encountered in the submitted
Libopt lines are valid (otherwise there is no verification). The
\texttt{\textit{list-of-performance-tokens}} is a blank-separated list
of tokens that can be considered as performance tokens. If the
directive \texttt{tokens} is present, it is verified that the
performance tokens are among the tokens in the
\texttt{\textit{list-of-tokens}}. The performance tokens are used by
\texttt{perfopt} and \texttt{queryopt} to determine whether the token
proposed to make the comparison of solvers is a performance token.

The third directive, \texttt{data\_base}, specifies the database name
\texttt{\textit{DBFile}} that has to be used when this one is not
specified in the command line of some commands that need such a
database (\texttt{addopt}, \texttt{perfopt}, \ldots). Hence the name
given on the command line has priority over the name given in the
startup file. If neither the command line nor the startup file provide
a file name, the commands assume that the database is the file
\texttt{dtbopt}\index{file!dtbopt default database name@\texttt{dtbopt}
(default database name)|ibf} in the working directory. The actual name
of the database on disk may have the extension ``\texttt{.db}'' or may
be represented by two files with names having the extensions
``\texttt{.dir}'' and ``\texttt{.pag}''; this depends on the operating
system.

An example of \texttt{liboptrc} file can be found in the
directory~\texttt{doc} (under the name \texttt{liboptrc\_}\abr
\texttt{optim}). It may be useful in optimization.

\index{file!.liboptrc startup file@\texttt{.liboptrc} (startup file)|)}

\subsection{Lists}
\label{s:lists}

\index{list|(}

The behavior of the Libopt scripts depends on various lists, which are
used to control the consistency of the commands with respect to the
content of the Libopt hierarchy.
These lists are located in various
directories and, for clarity, they have the suffix ``\texttt{.lst}''.
One finds lists of solvers, of collections, and of problems
(\texttt{\SUBC.lst}):
\begin{quote}
\index{list!of collections}%
\texttt{%
\$LIBOPT\_DIR/collections/collections.lst}\\
\index{list!of problems}%
\texttt{%
\$LIBOPT\_DIR/collections/\COLL/\SUBC.lst}\\
\index{list!of solvers}%
\texttt{%
\$LIBOPT\_DIR/solvers/solvers.lst}\\
\index{list!of collections}%
\texttt{%
\$LIBOPT\_DIR/solvers/\SOLV/collections.lst}\\
\index{list!of problems}%
\texttt{%
\$LIBOPT\_DIR/solvers/\SOLV/\COLL/\SUBC.lst}
\end{quote}
A list is simply a sequence of strings separated by spaces
(\texttt{\char`\\s+}). {\em Comments\/}\index{list!comment in a
--}\label{comments-in-list} are possible; they start from the character
`\texttt{\#}' up to the end of the line. Comments are used to describe
the list and its elements.

\index{list|)}

\subsection{The \texttt{runopt} command}
\label{s:runopt}

\index{command (Libopt --)!runopt@\texttt{runopt}|(}

\subsubsection{Overview}
\label{s:runopt-overview}

The \texttt{runopt} command is probably the most complex of those
offered by the Libopt environment. It is used to run solvers on
problems (see section~\ref{s:runopt-basic} for an introduction).
Because of its genericity, this command cannot realize itself all the
details of this operation. Indeed, it cannot know all the features of
any possible solver and any possible collection of problems (even those
that are still not installed in the hierarchy!). For this reason,
\texttt{runopt} must delegate part of the operations to other
scripts/programs, some of them being written by the persons introducing
new solvers and collections into the Libopt environment. The operation
of running solvers on problems is actually decomposed in three levels
of programs.
\begin{list}{{\small$\bullet$}}{\topsep=2.0ex\parsep=0.5ex\itemsep=1.0ex
 \settowidth{\labelwidth}{{\small$\bullet$}}
 \labelsep=0.5em
 \leftmargin=\bulindent
}
\item
{\sf The Libopt level.}\nopagebreak\\[1.0ex]
The first level is formed of the operations implemented in the
\texttt{runopt} script itself. These operations are those that are
independent of any solver and any problem collection. The aim of this
level is to analyze the \texttt{runopt} command line and to decompose
it in a sequence of elementary operations of the form
\begin{equation}
\label{elementary-operation}
\mbox{run the solver \SOLV\ on the problem \PROB\ of the collection
\COLL.}
\end{equation}
There are as many elementary operations as there are combinations of
solvers, collections, and problems expressed by the \texttt{runopt}
command lines. To execute each elementary operation, \texttt{runopt}
calls the program \texttt{\SOLV\_\COLL} of the second level.

The \texttt{runopt} script is further described in
section~\ref{s:runopt-script} below. Note that it has not to be
modified when the Libopt environment is enriched with new solvers or
new collections of problems.

\item
{\sf The operating-system level.}\nopagebreak\\[1.0ex]
The second level is the one that takes in charge the elementary
operation \eqref{elementary-operation}. It is realized by the programs
\begin{quote}
\texttt{%
\$LIBOPT\_DIR/solvers/\SOLV/\COLL/\SOLV\_\COLL}
\end{quote}
These are actually Perl scripts in the standard distribution. There is
a third level since, by choice and for more flexibility and
readability, the \texttt{\SOLV\_}\abr \texttt{\COLL} programs are
restricted to the necessary operations {\em at the operating system
level}. However, unlike \texttt{runopt}, these operations depend here
on the solver \SOLV\ and the problem collection \COLL. Typical
operations consist in copying data files in the working directory,
running makefiles to get executable programs, launching the main
program, and removing data and result files from the working directory
on exit from the main program. For some collections of problems (like
CUTEr), some of these operations may also have been delegated to other
programs intimately incorporated into the collection.

More is said about the \texttt{\SOLV\_\COLL} programs in
section~\ref{s:solv-coll-script}. Note that such a program has to be
written when a new solver or a new collection of problems is installed
in the Libopt environment. It can often be adapted from an existing
one; for example all the \texttt{\SOLV\_cuter} programs only differ be
a few character strings.

\item
{\sf The main program level.}\nopagebreak\\[1.0ex]
Running a program on a problem does not only depend on operations done
at the operating system level. A main program has to be written, which
reads the data, calls the solver, and delivers diagnosis. It is at
this level that the {\em Libopt line\/}\index{Libopt line} is generally
written on the standard output. This is the part of the Libopt
environment that is the most interlinked with the solver and problem
structures. As a result, such a program has to be written for each new
solver-collection pair. This level is further described in
section~\ref{s:main}.
\end{list}

\subsubsection{The \texttt{runopt} script}
\label{s:runopt-script}

The general form of the \texttt{runopt} command is the following
\begin{quote}
\index{command (Libopt --)!runopt@\texttt{runopt}|ibf}
\texttt{%
\% runopt [-h] [-k] [-t] [-v] [\textit{CommandFile}]}
\end{quote}
where \texttt{\textit{CommandFile}} is a file that contains command
lines of the form
\begin{quote}
\texttt{%
\SOLV[.\TAG] \COLL[.\SUBC] [\textit{list-of-problems}]}
\end{quote}
in which \SOLV\ is some solver, \TAG\index{tag} is an optional string
tagging the solver name in the Libopt line (the usefulness of this
option is discussed in section~\ref{s:tag}), \COLL\ is some collection,
\SUBC\ is an optional subcollection of \COLL, and
\texttt{\textit{list-}}\abr \texttt{\textit{of-}}\abr
\texttt{\textit{problems}} is an optional blank-separated list of
problems of the subcollection. If there is no file of commands,
\texttt{runopt} reads the standard input. Therefore, if there is just a
single command line, it may be easier to feed the standard input of
\texttt{runopt}, using the Unix/Linux command
\texttt{echo}\index{command (Unix/Linux --)!echo@\texttt{echo}} like on
page~\pageref{first-runopt}.

For each command line, the list of problems that are actually tried to
be solved by \SOLV\ is established by \texttt{runopt} in the following
manner.
\begin{list}{}{\topsep=0.5ex\partopsep=0.5ex\parsep=0.5ex\itemsep=0.5ex
 \settowidth{\labelwidth}{9.}
 \labelsep=0.5em
 \leftmargin=\labelwidth
 \addtolength{\leftmargin}{\labelsep}
}
\item[1.]
If there is no subcollection specified in the command line (no
``\texttt{.\SUBC}''), \texttt{runopt} assumes either the
``\texttt{all}'' subcollection if the
\texttt{\textit{list-of-problems}} is nonempty or the
``\texttt{default}'' subcollection otherwise. The logic behind this
rule is to be non-restrictive when some problems are mentioned in the
command line (the
``\texttt{all}''\index{subcollection!all@\texttt{all}}\index{list!all.lst@\texttt{all.lst}}
subcollection is supposed to give all the problems of the collection),
and to limit the number of problems to consider when no problem is
specified (the
``\texttt{default}''\index{subcollection!default@\texttt{default}}\index{list!default.lst@\texttt{default.lst}}
subcollection is supposed to give a not too large subset of typical
problems of a huge collection).

\item[2.]
Once a subcollection, say \SUBC\texttt{\textit{2}}, has been
determined, either from the one given in the command line or from the
rule just mentioned, \texttt{runopt} looks for a file describing the
subcollection \SUBC\texttt{\textit{2}}. It searches in the following
order:
\begin{quote}
\texttt{\COLL.\SUBC\textit{2}.lst} in the working directory\\
\texttt{\$LIBOPT\_DIR/solvers/\SOLV/\COLL/\SUBC\textit{2}.lst}\\
\texttt{\$LIBOPT\_DIR/collections/\COLL/\SUBC\textit{2}.lst}
\end{quote}
The logic is the following. Since \texttt{runopt} first looks in the
working directory, the user is allowed to build a temporary list of
problems for a special purpose. Next, priority is given to the lists
the solver \SOLV\ has declared in connection with the collection \COLL,
since it is the solver that knows the type of problems it can solve. If
no file \texttt{\SUBC\textit{2}.}\abr \texttt{lst}\index{list!of
problems} is found in the above locations, the command line is ignored.

\item[3.]
\texttt{Runopt} takes the additional precaution of eliminating from
\texttt{\SUBC\textit{2}} the problems that are not soluble by \SOLV\
(i.e., those that are not in \texttt{\$LIBOPT\_DIR/}\abr
\texttt{solvers/}\abr \texttt{\SOLV/}\abr \texttt{\COLL/}\abr
\texttt{all.}\abr \texttt{lst}). Let \texttt{\SUBC\textit{3}} be the
resulting list.

\item[4.]
The final list is then, either \texttt{\SUBC\textit{3}} if there is no
\texttt{\textit{list-of-problems}} in the command line, or the
intersection of \texttt{\SUBC\textit{3}} and the
\texttt{\textit{list-of-problems}} if the latter is nonempty. Let
\texttt{\SUBC\textit{4}} be the final list.

\end{list}
Then \texttt{runopt} launches the script \texttt{\$LIBOPT\_}\abr
\texttt{DIR/}\abr \texttt{solvers/}\abr \texttt{\SOLV/}\abr
\texttt{\COLL/}\abr \texttt{\SOLV\_}\abr \texttt{\COLL} for each
problem \PROB\ of the list \texttt{\SUBC\textit{4}}.

\subsubsection{The \SOLV\_\COLL\ scripts}
\label{s:solv-coll-script}

A \texttt{\SOLV\_\COLL} script must exist for each solver \SOLV\ that
has been installed in the Libopt environment to run problems from the
collection \COLL. It must be placed in the directory
\texttt{\$LIBOPT\_}\abr \texttt{DIR/}\abr \texttt{solvers/}\abr
\texttt{\SOLV/}\abr \texttt{\COLL} and is launched by the
\texttt{runopt} command. It must accept the following command line
structure
\begin{quote}
\index{command (Libopt --)!solv coll@\texttt{\SOLV\_}\abr
\texttt{\COLL}|ibf}
\texttt{%
\% \SOLV\_\COLL\ [-k] [-t] [-v] \PROB}
\end{quote}
The options \texttt{-k}, \texttt{-t}, and \texttt{-v} are inherited
from the \texttt{runopt} command that launches \texttt{\SOLV\_\COLL}.
It is the \texttt{runopt} command that determines the problem  \PROB\
to solve (see section~\ref{s:runopt-script}). As discussed in
section~\ref{s:runopt-overview}, the \texttt{\SOLV\_}\abr
\texttt{\COLL} program has in charge the operations, at the operating
system level, for running the solver \SOLV\ on the problem \PROB\ of
the collection~\COLL.

When there is already a solver, say \texttt{\textit{solv0}}, that has
been installed to run problems from the collection \COLL, it is often
quite easy to get the correct \texttt{\SOLV\_\COLL} script. It is
indeed usually sufficient to copy
\begin{quote}
\texttt{%
\% cp \$LIBOPT\_DIR/solvers/\textit{solv0}/\COLL/\textit{solv0}\_\COLL \char`\\}\\
\hphantom{\texttt{\% cp }}\texttt{%
\$LIBOPT\_DIR/solvers/\SOLV/\COLL/\SOLV\_\COLL}
\end{quote}
and to substitute in the copied file every instance of the string
``\texttt{\textit{solv0}}'' by the string ``\SOLV''. Examples exist in
the standard distribution, which have this property due to a high
degree of parametrization of the scripts.


Consider now the general case, when one has to start from scratch,
because a new collection \COLL\ is installed in the hierarchy. Let us
describe the operations that the script \texttt{\SOLV\_\COLL} must
realize, for a particular solver \SOLV. The problem to solve, \PROB, is
given by the single plain argument of the command. We have said above
that the script has to take care of the options \texttt{-k},
\texttt{-t}, and \texttt{-v} of the command. Next, the script
\texttt{\SOLV\_\COLL} must realize three tasks.
\begin{list}{}{\topsep=0.5ex\partopsep=0.5ex\parsep=0.5ex\itemsep=0.5ex
 \settowidth{\labelwidth}{9.}
 \labelsep=0.5em
 \leftmargin=\labelwidth
 \addtolength{\leftmargin}{\labelsep}
}
\item[1.]
Build in the working directory the executable program, say
\texttt{\SOLV\_}\abr \texttt{\COLL\_}\abr \texttt{main} (a binary code
that will solve the problem \PROB\ using the solver \SOLV), and copy
into the working directory, from somewhere in the hierarchy
\texttt{\$LIBOPT\_}\abr \texttt{DIR/}\abr \texttt{collections/}\abr
\texttt{\COLL}, the data files that are useful for running
\texttt{\SOLV\_}\abr \texttt{\COLL\_}\abr \texttt{main}.

These operations depend in depth on the installation of the collection
\COLL.
\begin{list}{{\small$\bullet$}}{\topsep=1.5ex\parsep=0.5ex\itemsep=0.5ex
 \settowidth{\labelwidth}{{\small$\bullet$}}
 \labelsep=0.5em
 \leftmargin=\labelwidth
 \addtolength{\leftmargin}{\labelsep}
}
\item
For the Modulopt\index{collection!Modulopt} collection, the operations
are encoded at appropriate targets in the makefiles:
\begin{quote}
\texttt{%
\$LIBOPT\_DIR/collections/modulopt/probs/\PROB/Makefile},\\
\texttt{%
\$LIBOPT\_DIR/solvers/\SOLV/modulopt/Makefile}.
\end{quote}
Therefore, \texttt{\SOLV\_}\abr \texttt{\COLL} just execute these
targets after having set the two environment variables used by them,
which specify the problem name (\texttt{\$MODULOPT\_PROB}) and the
working directory (\texttt{\$WORKING\_DIR}).

\item
For the CUTEr\index{collection!CUTEr} collection, this first task could
have been skipped, since the CUTEr command \texttt{sd\SOLV} (or
something similar) takes it in charge. However, for allowing to store
the problems in a compressed format, the file \texttt{\PROB.}\abr
\texttt{SIF.}\abr \texttt{gz} is copied in the working directory and 
decompressed there.
\end{list}

\item[2.]
Run the executable program \texttt{\SOLV\_}\abr \texttt{\COLL\_}\abr
\texttt{main} (or \texttt{sd\SOLV} for the CUTEr collection) in the
working directory.

\item[3.]
Remove from the working directory the now useless executable program
\texttt{\SOLV\_}\abr \texttt{\COLL\_}\abr \texttt{main} (not necessary
for the CUTEr collection), the data files, and the output files that
have been generated during the execution of the program. This task also
depends on the collection \COLL\ and the solver \SOLV. For the CUTEr
collection, part of the task is taken in charge by the command
\texttt{sd\SOLV} itself.
\end{list}

\subsubsection{The main program}
\label{s:main}

This is the main program that would have to be written if one has to
run the solver \SOLV\ on a particular problem \PROB\ of the collection
\COLL. The required genericity of this main program (it has to be able
to consider {\em any\/} problem \PROB\ of the collection) is obtained
in the following way. All the problems of the collection are described
by procedures with a name that depends on the realized function but
{\em is identical for all the problems}. Typically, one finds
\begin{list}{{\small$\bullet$}}{\topsep=1.5ex\parsep=0.0ex\itemsep=0.0ex
 \settowidth{\labelwidth}{{\small$\bullet$}}
 \labelsep=0.5em
 \leftmargin=\bulindent
}
\item
{\em initialization procedures}, which specify the dimensions of the
problem, possibly its name, initialize the variables, read the data,
etc,
\item
a {\em simulator}, which gives the state of the system to solve at a
given iterate when the problem is nonlinear and the solver has an
iterative nature,
\item
{\em auxiliary procedures}, which precise the way some objects are
computed (for example, the procedure that realizes the inner product
used for computing a gradient),
\item
and possibly {\em post-solution procedures}, which can realize some
post-solution computations.
\end{list}
The connection between the main program and the problem is then
realized by the link editor, to which the object files associated with
the main program and the procedures describing the selected problem are
given by the \texttt{\SOLV\_}\abr \texttt{\COLL} script described in
section~\ref{s:solv-coll-script}.

It is normally the main program that prints the {\em Libopt
line\/}\index{Libopt line} on the standard output. It can do this after
the problem has been solved by collecting the various features of the
problem and the various counters that depict the behavior of the
solver \SOLV\ on the problem \PROB.

\index{command (Libopt --)!runopt@\texttt{runopt}|)}

\subsection{The \texttt{addopt} command}
\label{s:addopt}

\index{command (Libopt --)!addopt@\texttt{addopt}|(}

We have discussed in large in section~\ref{s:addopt-basic}, the role of
the \texttt{addopt} command in the basic work cycle of the Libopt
environment. In particular, all the details on the Libopt line have
been presented in that section. In the present section, we make it
precise the form of the \texttt{addopt} command line.

The accepted forms of the \texttt{addopt} command are the following
\begin{quote}
\texttt{\% addopt [-h] [-t] [-v] [-b \textit{DBFile}]
[-r] \textit{ResFile}}
\\
\texttt{\% addopt [-h] [-t] [-v] [-b \textit{DBFile}]
-d \textit{selection}}
\end{quote}
If the option \texttt{-h} is present, \texttt{addopt} prints a short
help message describing the command usage and exits. The first command
is used for adding results in a database, the second for deleting
results.

The database \texttt{\textit{DBFile}} is specified with the \texttt{-b}
option. If this option is not present, \texttt{addopt} looks for the
database specified by the directive \texttt{data\_base} in the startup
file \texttt{\char`~/.liboptrc}. Finally, if this one does not exist,
\texttt{addopt} assumes that the database is the file
\texttt{dtbopt}\index{file!dtbopt default database name@\texttt{dtbopt}
(default database name)} in the working directory (see
section~\ref{s:liboptrc} for the details). The file \texttt{ResFile}
used in the first form of the command is supposed to contain Libopt
lines\index{Libopt line}, typically generated by the \texttt{runopt}
command. A result corresponding to a triple (solver, collection,
problem), already existing in the database is not replaced, unless the
option~\texttt{-r} is present.

The string \texttt{selection} used in the second form is interpreted as
a file name if it is not terminated by the character `\texttt{\%}' or
as a way of selecting triples (solver, collection, problem) otherwise.
If \texttt{selection} is view as a file name, this file is supposed to
be formed of Libopt lines and the results with the same triples
(solver, collection, problem) as in the Libopt lines are deleted from
the database. If \texttt{selection} is the string
``\texttt{\SOLV\%}\abr \texttt{\COLL\%}\abr \texttt{\PROB\%}'', the
result corresponding to the triple (solver, collection, problem) =
(\SOLV, \COLL, \PROB) is deleted from the database (if present). An
empty solver (resp.\ collection, problem) field in the string
\texttt{selection} matches any solver (resp.\ collection, problem).
For instance, if \texttt{selection} is the string
``\texttt{\SOLV\%\%\%}'' or simply ``\texttt{\SOLV\%}'', all the
results corresponding to the solver \SOLV\ are deleted from the
database. As another example, if \texttt{selection} is the string
``\texttt{\%\COLL\%\PROB\%}'', all the results corresponding to the
problem \PROB\ of the collection \COLL\ are deleted from the database.
The option \texttt{-t} can be used to see the effect that a deleting
command would have before really doing~it.

\index{command (Libopt --)!addopt@\texttt{addopt}|)}

\subsection{The \texttt{perfopt} command}
\label{s:perfopt}

\index{command (Libopt --)!perfopt@\texttt{perfopt}|(}

The \texttt{perfopt} command was introduced in
section~\ref{s:comparing-results}, where its role in the basic work
cycle of the Libopt environment was shown. In this section, we describe
the command in detail. It has the following form:
\begin{quote}
\index{command (Libopt --)!perfopt@\texttt{perfopt}|ibf}
\texttt{\% perfopt [-h] [-v] [-b \textit{DBFile}] [-p \textit{ptok}]
[-log] [-g \textit{GFile}]}
\end{quote}
The script is used to compare the performance of various solvers on a
given set of problems. A single criterion is used for the comparison
and it can be specified by the option ``\texttt{-p}
\texttt{\textit{ptok}}'',
where \texttt{\textit{ptok}} is one of the {\em performance
tokens\/}\index{token!performance} in the Libopt line (see
sections~\ref{s:addopt-basic} and~\ref{s:liboptrc}); another way of
specifying this performance criterion is to use the specification file
\texttt{perfopt.}\abr \texttt{spc} (see below). The results on which
the considered solvers are compared are supposed to be stored in the
database \textit{\texttt{DBFile}} (if the \texttt{-b} option is not
used, \texttt{perfopt} uses the name given in the startup file or the
default name \texttt{dtbopt}\index{file!dtbopt default database
name@\texttt{dtbopt} (default database name)}, see
section~\ref{s:liboptrc} for the details), typically built by the
\texttt{addopt} command. The \texttt{perfopt} command generates two
files: \textit{\texttt{GFile}}\abr \texttt{.m} and
\textit{\texttt{GFile}}\abr \texttt{.gnu} (the names
\texttt{perf.m}\index{file!perf.m@\texttt{perf.m}} and
\texttt{perf.gnu}\index{file!perf.gnu@\texttt{perf.gnu}} are used if
the option \texttt{-b} is not present in the command line). The file
\textit{\texttt{GFile}}\abr \texttt{.m} is a Matlab function providing
a graphic representation of the performance profiles of the compared
solvers, \textit{\texttt{GFile}}\abr \texttt{.gnu} is a Gnuplot data
file of the same performance profiles. The option \texttt{-log}
requires that the performance profiles must have a logarithmic
x-coordinate.

\index{directive!perfopt@\texttt{perfopt}|(}

The \texttt{perfopt} command is best tuned by using {\em directives\/}
in the specification file \texttt{perfopt.spc}\index{file!perfopt
spc@\texttt{perfopt.}\abr \texttt{spc} (specification file)} in the
working directory. When starting, \texttt{perfopt} tries to read this
file  and terminates if it cannot find and read it. Therefore, {\em
this file is mandatory}. The file is described in the manual page of
\texttt{perfopt}. To make the discussion that follows understandable,
some of the directives of the specification file have to be clarified.
First, at least two solvers have to be selected for the comparison, say
\SOLVa\ and \SOLVb, which is done by one or more directives of the form
\begin{quote}
\texttt{solver \SOLVa\ \SOLVb\ ...}
\end{quote}
This is because \texttt{perfopt} cannot choose default solvers and it
makes no sense to compare a solver  with itself. Optionally, some
problems may have been selected by one or more directives of the form
\begin{quote}
\texttt{collection \COLL[.\SUBC] ...}
\end{quote}
The list of problems specified by this directive is searched in order
in the following files
\begin{quote}
\texttt{\COLL.\SUBC.lst} in the working directory\\
\texttt{\$LIBOPT\_DIR/collections/\COLL/\SUBC.lst}
\end{quote}
where \SUBC\ is set to the string `\texttt{all}' if it is not specified
by the directive. Finally, a performance token must have been
specified, say \textit{\texttt{ptok}}, either in the command line of
\texttt{perfopt} or by a directive of the form
\begin{quote}
\texttt{performance \textit{ptok}}
\end{quote}

\index{directive!perfopt@\texttt{perfopt}|)}%

Here is how \texttt{perfopt} selects the actual problems that will be
used for the comparison of the solvers specified by the
`\texttt{solver}' directives. These problems are those that satisfy the
following conditions:
\begin{list}{{\small$\bullet$}}{\topsep=1.5ex\parsep=0.0ex\itemsep=0.0ex
 \settowidth{\labelwidth}{{\small$\bullet$}}
 \labelsep=0.5em
 \leftmargin=\bulindent
}
\item
they are among those problems that have been specified by a
`\texttt{collec}\-\texttt{tion}' directive, if any such directive as
been used in the specification file \texttt{perfopt}\abr \texttt{.spc},
\item
they are present in the database \texttt{\textit{DBFile}} with a
\texttt{\textit{ptok}} performance token,
\item
they have not been discarded by the `\texttt{problem}' directive (see
the manual page of \texttt{perfopt}),
\item
they have been solved by all the solvers specified by the
`\texttt{solver}' directives.
\end{list}

\index{command (Libopt --)!perfopt@\texttt{perfopt}|)}

\subsection{Getting results from a modified version of an installed
solver}
\label{s:tag}

Suppose that some parameters of some solver, say \texttt{goya}, are
modified and that it is desirable to see the effect of these
modifications on the performances of the solver. This situation also
occurs during the development of a solver, when it is desirable to see
whether a newly developed technique has a good effect, by comparing the
results of the new version with the previous ones. We see three ways of
realizing this and consider them from the hardest to the easiest one.

\begin{list}{}{\topsep=0.5ex\partopsep=0.5ex\parsep=0.5ex\itemsep=0.5ex
 \settowidth{\labelwidth}{9.}
 \labelsep=0.5em
 \leftmargin=\labelwidth
 \addtolength{\leftmargin}{\labelsep}
}
\item[1.]
A first possibility would be to define a full new hierarchy under the
\texttt{\$LIBOPT\_}\abr \texttt{DIR/}\abr \texttt{solvers} directory
with the new version of the solver. This results in defining a new row
in the $\{$solvers$\} \times \{$collections$\}$ Cartesian product. This
is certainly safe, but it is not an easy and rapid task. In addition,
the solver with the new technique may not exist for a long time.
Therefore, this option is probably not the best idea.

\item[2.]
A better way of doing is to modify temporarily the code that generates
the Libopt line\index{Libopt line}, so that instead of generating a
line of the form
\begin{quote}
\begin{verbatim}
libopt%goya.quatro-de-junio% ...
\end{verbatim}
\end{quote}
where \texttt{goya} has been changed into \texttt{goya.}\abr
\texttt{quatro-}\abr \texttt{de-}\abr \texttt{junio}. The modified
Libopt lines can also be obtained by editing the file containing the
Libopt lines of the new version of the solver. The \texttt{addopt}
command will believe that this line has been generated by the solver
\texttt{goya.}\abr \texttt{quatro-}\abr \texttt{de-}\abr
\texttt{junio}, even though judiciously this one does not exist in the
Libopt hierarchy. Therefore, the comparison will be made without
difficulty with other solvers whose results have been stored in some
database (using the command \texttt{addopt}), possibly with
\texttt{goya.}\abr \texttt{tres-}\abr \texttt{de-}\abr \texttt{mayo},
the presumably best version of \texttt{goya} obtained so~far.

\item[3.]
Alternatively, instead of modifyng the code generating the Libopt
lines, one could modify these lines after they have been written in a
file, using a text editor. Libopt offers an automatic way of getting
the same effect. It is indeed possible to add a {\em
tag\/}\index{tag|ibf} to the solver name in the Libopt line. This is
obtained by entering
\begin{quote}%
\texttt{%
\% echo "goya.quatro-de-junio \COLL\ \PROB" | runopt}%
\index{command (Unix/Linux --)!echo@\texttt{echo}}%
\index{command (Libopt --)!runopt@\texttt{runopt}}%
\end{quote}%
Then, \texttt{runopt} concatenates the string
``\texttt{.quatro-de-junio}'' to the solver name in the Libopt
line\index{Libopt line} of the standard output, which then becomes
\begin{quote}
\texttt{%
libopt\%goya.quatro-de-junio\%\COLL\%\PROB\%\textit{...}}
\end{quote}
Of course, \texttt{runopt} considers that the solver is \texttt{goya}
not \texttt{goya.}\abr \texttt{quatro-}\abr \texttt{de-}\abr
\texttt{junio} (because of the dot `\texttt{.}', the latter does not
exist in the Libopt hierarchy). This effect is obtained by filtering
the standard output of \texttt{runopt} with the stream editor
\texttt{sed}\index{command (Unix/Linux --)!sed@\texttt{sed}}, which
only modifies the Libopt lines\index{Libopt line}.

\end{list}

\begin{figure}[htbp]
\centering
\psfrag{m1qn3.diag}{\footnotesize\texttt{{ diag}}}
\psfrag{m1qn3.scal}{\footnotesize\texttt{{ scal}}}
\includegraphics[scale=0.5]{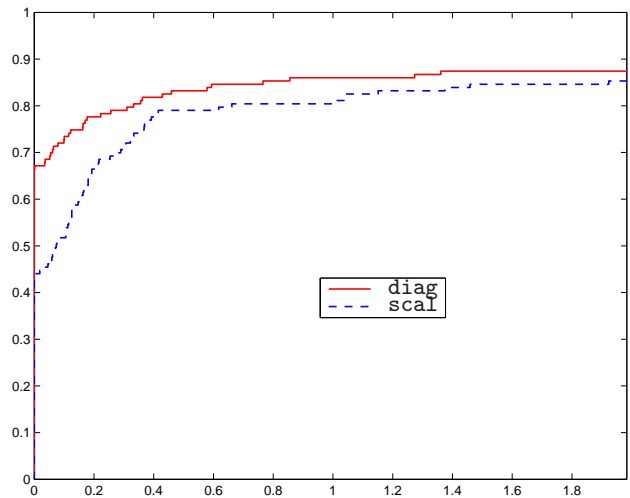}
\caption{\label{f:m1qn3-diag-scal}Performance profiles of the diagonal
(\texttt{diag}, red solid curve) and scalar (\texttt{scal}, blue dashed
curve) running modes of \texttt{m1qn3} on 143 unconstrained problems of
the CUTEr collection, comparing the number of function evaluations}
\end{figure}

For example, if it is desired to compare the scalar and diagonal
running modes of the $\ell$-BFGS code
\texttt{m1qn3}~\cite{gilbert-lemarechal-1989} on 143 unconstrained
CUTEr problems, one can proceed as follows. Run \texttt{m1qn3} with its
default setting (diagonal running mode), tagging the solver with the
string ``\texttt{diag}'' to remember the option used in the run
\begin{quote}
\index{command (Libopt --)!runopt@\texttt{runopt}}
\texttt{%
\% echo "m1qn3.diag cuter.unc" | runopt | grep libopt
\char`\\}\\
\hphantom{\texttt{\% }}\texttt{%
> m1qn3.diag-cuter.unc.lbt}
\end{quote}
Set the option \texttt{imode(1) = 1} in the main program
\texttt{m1qn3ma.f} used by the CUTEr hierarchy to put \texttt{m1qn3} in
scalar running mode, generate the object file \texttt{m1qn3ma.o}, put
it in the appropriate place in the CUTEr hierarchy, and run again:
\begin{quote}
\index{command (Libopt --)!runopt@\texttt{runopt}}
\texttt{%
\% echo "m1qn3.scal cuter.unc" | runopt | grep libopt
\char`\\}\\
\hphantom{\texttt{\% }}\texttt{%
> m1qn3.scal-cuter.unc.lbt}
\end{quote}
Add the results obtained in this way in the database:
\begin{quote}
\texttt{\% addopt m1qn3.diag-cuter.unc.lbt}\\
\texttt{\% addopt m1qn3.scal-cuter.unc.lbt}
\end{quote}
Adapt the \texttt{solver} directive of the specification file
\texttt{perfopt.spc}:
\begin{quote}
\texttt{solver m1qn3.diag m1qn3.scal}
\end{quote}
Now enter
\begin{quote}
\texttt{\% perfopt}
\end{quote}
This last command generates the file ``\texttt{perf.m}'', a Matlab
M-file, which can be used to plot the performance profiles of the two
versions of \texttt{m1qn3}, those that are given in
figure~\ref{f:m1qn3-diag-scal}.
An accustomed reader can deduced from these curves that the diagonal
mode of \texttt{m1qn3} can be considered as slightly more efficient
than the scalar mode (its profile is higher, see
section~\ref{s:comparing-results}). This is an average estimation,
since the plot with its $2$-logarithmic abscissa reveals that, if the
scalar mode can require $2^{1.9}\simeq3.7$ times more function
evaluations than the diagonal mode, the latter can also require
sometimes $2^{1.4}\simeq2.6$ times more evaluations than the former.

\subsection{Managing platforms}
\label{s:platforms}

\index{platform|(}

The environment variable \texttt{\$LIBOPT\_}\abr
\texttt{PLAT}\index{environment variable!LIBOPT
PLAT@\texttt{LIBOPT\_}\abr \texttt{PLAT}} provides the {\em
platform\/}\index{platform} on which the user works. This is a string
of the form ``\MACH.\OS.\COMP'', where \MACH\ designates the machine,
\OS\ is its operating system, and \COMP\ is the compiler suite. The
current standard distribution of Libopt includes the following
platforms:
\begin{quote}
\index{mac osx gcc platform@\texttt{mac.osx.gcc} (platform)}%
\texttt{mac.osx.gcc}\\
\index{pc lnx pg platform@\texttt{pc.lnx.pg} (platform)}%
\texttt{pc.lnx.pg}
\end{quote}
where
\begin{list}{}{\topsep=1.5ex\partopsep=0.5ex\parsep=0.0ex\itemsep=0.0ex
 \settowidth{\labelwidth}{\texttt{xxx}}
 \labelsep=1.0em
 \leftmargin=\labelwidth
 \addtolength{\leftmargin}{\labelsep}
 \addtolength{\leftmargin}{2.5em}
}
\item[\texttt{mac}\hfill]
Macintosh computer (Apple),
\item[\texttt{pc}\hfill]
PC-like computer,
\vspace{1ex}
\item[\texttt{lnx}\hfill]
Linux operating system,
\item[\texttt{osx}\hfill]
OSX operating system (Macintosh),
\vspace{1ex}
\item[\texttt{gcc}\hfill]
GNU Compiler Collection,
\item[\texttt{pg}\hfill]
Portland Group compilers.
\end{list}

The platform name is used by some scripts and makefiles of the Libopt
environment. The Libopt hierarchy of the standard distribution contains
indeed no object/compiled files. When such files are required they are
generated in the working directory and the value of
\texttt{\$LIBOPT\_}\abr \texttt{PLAT}\index{environment variable!LIBOPT
PLAT@\texttt{LIBOPT\_}\abr \texttt{PLAT}} is used to generate them.
This description is given by files of the form
\begin{quote}
\texttt{%
make.\textit{plat}}
\end{quote}
where ``\texttt{\textit{plat}}'' is one of the strings above.
These files are located in the directory
\begin{quote}
\texttt{%
\$LIBOPT\_DIR/platforms}
\end{quote}
A new platform is therefore obtained by adding a new file
\texttt{make.\textit{plat}} in that directory, adapting the various
variables defined in the file. It is used by resetting the environment
variable \texttt{\$LIBOPT\_}\abr \texttt{PLAT}.

\index{platform|)}

\section{Hooking a collection of problems onto Libopt}
\label{s:adding-collection}

There are probably many ways of hooking a collection of problems to the
Libopt environment. This installation depends in great part on the
nature of the considered collection. For example, the installation of
the CUTEr and Modulopt collections are quite different. Almost
everything is possible provided the scripts using this collection
reflect the options taken during the installation. This is why Libopt
can just be viewed as an empty shell (the directory structure) and a
methodology (embodied in its scripts).

The installation of a collection is usually quite simple. Two typical
installations are those of the CUTEr and Modulopt collections. We
consider them in sequence.

\subsection{Hooking the CUTEr collection}
\label{s:hooking-cuter-collection}

\index{collection!CUTEr|(}

Libopt is a layer covering CUTEr. Therefore, CUTEr must already be
installed before being hooked onto Libopt. On the other hand, the
standard distribution of Libopt provides part of the installation of
CUTEr into Libopt, since the directory \texttt{\$LIBOPT\_}\abr
\texttt{DIR/}\abr \texttt{collections/}\abr \texttt{cuter}
exists and contains some lists of problems \texttt{*.lst}. Actually,
the only thing that the installed scripts using CUTEr require to be
done is to define
\begin{quote}
\texttt{\$LIBOPT\_DIR/collections/cuter/probs}
\end{quote}
as a symbolic link to the directory containing the problems of the
CUTEr collection (files with the suffix ``\texttt{.SIF}''), usually
denoted by the environment variable
\texttt{\$MASTSIF}\index{environment
variable!MASTSIF@\texttt{MASTSIF}}. The scripts present in the standard
distribution of Libopt assume that these files are stored in
uncompressed or \texttt{gzip}-compressed format.

The installation of the CUTEr collection is therefore quite simple.
This does not mean that it can be readily used. Indeed, the solvers
installed in Libopt have now to be prepared to run CUTEr problems. This
has already been achieved for the solvers in the standard distribution,
but some adjustments may be necessary for a new solver; this is
discussed in section~\ref{s:solver-cuter}.

\index{collection!CUTEr|)}

\subsection{Hooking a new collection}
\label{s:hooking-toys-collection}

In this section, we discuss the installation in the Libopt environment
of a typical collection of problems. Let us call it
\begin{quote}
\texttt{toys}
\end{quote}
The installation is modeled on the one of the Modulopt collection,
already present in the standard distribution of
Libopt~\cite{gilbert-2007b}. We consider first the definition of the
collection into Libopt and next a possible way of introducing problems
into the collection.

\subsubsection{Defining a new collection in the Libopt environment}

\begin{list}{}{\topsep=0.5ex\partopsep=0.5ex\parsep=0.5ex\itemsep=0.5ex
 \settowidth{\labelwidth}{9.}
 \labelsep=0.5em
 \leftmargin=\labelwidth
 \addtolength{\leftmargin}{\labelsep}
}
\item[1.]
Define the new directories
\begin{quote}
\texttt{%
\$LIBOPT\_DIR/collections/toys}\\
\texttt{%
\$LIBOPT\_DIR/collections/toys/probs}
\end{quote}
and add the name \texttt{toys} to the list
\begin{quote}
\texttt{%
\$LIBOPT\_DIR/collections/collections.lst}
\end{quote}
Put all the problems of the collection \texttt{toys} in \texttt{probs}
(alternatively, you can make \texttt{probs} a symbolic link to the
directory containing all the problems of the collection). Each problem
can be stored in a dedicated directory within \texttt{probs}.

\item[2.]
Create the files
\begin{quote}
\texttt{%
\$LIBOPT\_DIR/collections/toys/all.lst}\\
\texttt{%
\$LIBOPT\_DIR/collections/toys/default.lst}
\end{quote}
The file \texttt{all.lst}\index{list!all.lst@\texttt{all.lst}}
lists all the problems of the collection \texttt{toys}. The file
\texttt{default.}\abr
\texttt{lst}\index{list!default.lst@\texttt{default.lst}} lists a
subset of the problems of the collection; this list is chosen by some
scripts when no list is specified as one of their arguments; therefore,
it can be a symbolic link to the list \texttt{all.}\abr \texttt{lst},
obtained using the Unix/Linux command `\texttt{ln} \texttt{-s}
\texttt{all.lst} \texttt{default.}\abr \texttt{lst}'; however, if the
collection contains many problems, it is probably better the put in
\texttt{default.lst} a not too long list of problems reflecting the
collection. Other lists of problems can also be defined in the
directory \texttt{toys}.

All these files may be empty is the collection is started from scratch.

\end{list}

\subsubsection{Introducing a new problem in the \texttt{toys}
collection}
\label{s:introducing-prob-in-toys}

We assume below that the each problem of the \texttt{toys} collection
is described in the dedicated directory
\begin{quote}
\texttt{%
\$LIBOPT\_DIR/collections/toys/probs/\PROB}.
\end{quote}
Hence, let \PROB\ be the name of the new problem we want to introduce.
This can be done as follows.

\begin{list}{}{\topsep=0.5ex\partopsep=0.5ex\parsep=0.5ex\itemsep=0.5ex
 \settowidth{\labelwidth}{9.}
 \labelsep=0.5em
 \leftmargin=\labelwidth
 \addtolength{\leftmargin}{\labelsep}
}
\item[1.]
Insert the name \PROB\ in the list of \texttt{toys} problems
\begin{quote}
\begin{verbatim}
$LIBOPT_DIR/collections/toys/all.lst
\end{verbatim}
\end{quote}
(and possibly in other lists in the same directory). This should be an
\texttt{ascii} file and {\em comments\/} should be possible (they
should start from the character `\texttt{\#}' up to the end of the
line).

\item[2.]
Create the directory
\begin{quote}
\texttt{%
\$LIBOPT\_DIR/collections/toys/probs/\textit{prob}},
\end{quote}
and put in that directory, all the files that define the problem \PROB:
source files, header files (if appropriate), and data files (if
appropriate).

\item[3.]
Create the makefile
\begin{quote}
\texttt{%
\$LIBOPT\_DIR/collections/toys/probs/\textit{prob}/Makefile},
\end{quote}
with the following two targets:
\begin{list}{{\small$\circ$}}{\topsep=1.0ex\parsep=0.5ex\itemsep=0.5ex
 \settowidth{\labelwidth}{{\small$\circ$}}
 \labelsep=0.5em
 \leftmargin=\labelwidth
 \addtolength{\leftmargin}{\labelsep}
 \addtolength{\leftmargin}{\parindent}
 \rightmargin=\parindent
}
\item
\PROB, which specifies how to obtain in the working directory an
archive \texttt{\textit{prob}.a} with all the object files defining
\PROB\ and which copies into the working directory the data files
needed for solving the problem (making symbolic links is probably
faster);
\item
\texttt{\textit{prob}\_clean}, which specifies which files has to be
removed from the working directory after having solved \PROB.
\end{list}
\end{list}

\section{Installing a solver into Libopt}
\label{s:solver}

We have seen in section~\ref{s:adding-collection} that the installation
of a new collection in Libopt is quite simple. However, just as it is,
no solver installed in the Libopt environment can use the new
collection \texttt{toys}, since the cells of the column
\texttt{solvers}$ \times \{$\texttt{toys}$\}$ of the Cartesian product
\eqref{cartesian-product} have not been filled in (here
\texttt{solvers} denotes the set of installed solvers, while
$\{$\texttt{toys}$\}$ denotes the singleton made of the single
collection \texttt{toys}). We consider the following possibilities.
\begin{list}{{\small$\bullet$}}{\topsep=1.5ex\parsep=0.5ex\itemsep=0.5ex
 \settowidth{\labelwidth}{{\small$\bullet$}}
 \labelsep=0.5em
 \leftmargin=\bulindent
}
\item
The solver does not exist in the Libopt environment: then follow the
instructions in section~\ref{s:solver-libopt} and consider the other
possibilities next.
\item
The solver exists in the Libopt environment and it is desired that it
solves problems from the CUTEr collection: see
section~\ref{s:solver-cuter}.
\item
The solver exists in the Libopt environment and it is desired that it
solves problems from an installed collection different from CUTEr: see
section \ref{s:interfacing-baby-toys}, where the Modulopt collection is
considered as a generic case. If the collection is not yet installed,
consider first section~\ref{s:adding-collection}.
\end{list}

\noindent
Throughout the section, we assume that
\begin{quote}
\texttt{baby}
\end{quote}
is the name of the considered solver.

\subsection{Defining a new solver in the Libopt environment}
\label{s:solver-libopt}

We gather in this section the operations that must be done for defining
the new solver \texttt{baby} in the Libopt environment, independently
of any specific collection. Here they~are.
\begin{list}{}{\topsep=0.5ex\partopsep=0.5ex\parsep=0.5ex\itemsep=0.5ex
 \settowidth{\labelwidth}{9.}
 \labelsep=0.5em
 \leftmargin=\labelwidth
 \addtolength{\leftmargin}{\labelsep}
}
\item[1.]
The name \texttt{baby} has to be added to the file
\begin{quote}
\texttt{%
\$LIBOPT\_DIR/solvers/solvers.lst},
\end{quote}
which contains the list of solvers of the Libopt environment. This file
is considered by Libopt as list of strings separated by
`\texttt{\char`\\s+}'.

\item[2.]
The following directories must be created
\begin{quote}
\texttt{%
\$LIBOPT\_DIR/solvers/baby}\\
\texttt{%
\$LIBOPT\_DIR/solvers/baby/bin}.
\end{quote}
The directory \texttt{baby} is the one that will contain the
description of all the collections to which the solver \texttt{baby}
will have access. The directory \texttt{bin} is a possible location for
the binaries and libraries defining the solver (or symbolic links to
these).

\item[3.]
Create also an empty file
\begin{quote}
\texttt{%
\$LIBOPT\_DIR/solvers/baby/collections.lst},
\end{quote}
which contains the list of collections that the solver \texttt{baby}
can consider (none if this is the first time \texttt{baby} is
introduced in the Libopt environment).

\item[4.]
It is recommended to write a file, named \texttt{README\_}\abr
\texttt{install}, that explains how the scripts that you are going to
introduce latter in the Libopt hierarchy, under the directory
\begin{quote}
\texttt{%
\$LIBOPT\_DIR/solvers/baby}
\end{quote}
assume to have access to the libraries and binaries
of the solver.
\end{list}

That is all for the installation of the solver \texttt{baby} itself.
Now, this solver cannot access any collection. For this, follow the
instructions in section~\ref{s:solver-cuter} for connecting the solver
\texttt{baby} to the CUTEr collection or in
section~\ref{s:interfacing-baby-toys} for another type of collection.

\subsection{Interfacing a solver with the CUTEr collection}
\label{s:solver-cuter}

We suppose here that
\begin{list}{{\small$\bullet$}}{\topsep=1.5ex\parsep=0.0ex\itemsep=0.0ex
 \settowidth{\labelwidth}{{\small$\bullet$}}
 \labelsep=0.5em
 \leftmargin=\bulindent
}
\item
the solver \texttt{baby} has already been
installed in the CUTEr environment (for more information, see the
manual~\cite{gould-orban-toint-2003}). This implies, in particular,
that a driver or main program \texttt{babyma} has been inserted in the
CUTEr environment and that the command \texttt{sdbaby} can be used, as
in
\begin{quote}
\texttt{%
\% sdbaby \PROB},
\end{quote}
where \PROB\ is the name of a problem in the CUTEr collection.

\item
the CUTEr collection has been installed in the Libopt environment (see
section~\ref{s:hooking-cuter-collection}),

\item
the solver \texttt{baby} has been introduced in the Libopt environment
(see section~\ref{s:solver-libopt}).
\end{list}

We are now ready to establish the connection between the solver
\texttt{baby} and the CUTEr collection, within Libopt. We recommend the
following steps.
\begin{list}{}{\topsep=0.5ex\partopsep=0.5ex\parsep=0.5ex\itemsep=0.5ex
 \settowidth{\labelwidth}{9.}
 \labelsep=0.5em
 \leftmargin=\labelwidth
 \addtolength{\leftmargin}{\labelsep}
}
\item[1.]
Create the directory
\begin{quote}
\texttt{%
\$LIBOPT\_DIR/solvers/baby/cuter}
\end{quote}
and add the name `\texttt{cuter}' to the list of collections that
\texttt{baby} can solve, namely
\begin{quote}
\texttt{%
\$LIBOPT\_DIR/solvers/baby/collections.lst}.
\end{quote}

\item[2.]
Define the script
\begin{quote}
\texttt{%
\$LIBOPT\_DIR/solvers/baby/cuter/baby\_cuter}.
\end{quote}
The easiest way of doing this is to copy and modify the corresponding
script of another solver, say \SOLV:
\begin{quote}
\texttt{%
\% cp \$LIBOPT\_DIR/solvers/\SOLV/cuter/\SOLV\_cuter}\\
\hphantom{\texttt{\% cp }}\texttt{%
\$LIBOPT\_DIR/solvers/baby/cuter/baby\_cuter}
\end{quote}
The modifications to bring to the content of the file
\texttt{baby\_cuter} are quite simple: change all the instances of
\SOLV\ into \texttt{baby} (you may want to be careful with some
capitalized letters in comment lines). An adaptation of the very last
lines of the file \texttt{baby\_}\abr \texttt{cuter}, dealing with the
removal of the files generated by the solver \texttt{baby} may require
some personal Perl writing (you can just forget this aspect of the
script if you prefer, nothing dangerous will occur).

\item[3.]
Modify the executable main program `\texttt{babyma}' installed in the
CUTEr environment, so that it prints the Libopt line, defined
in section~\ref{s:addopt-basic} (it is also defined in the manual page
`\texttt{libopt}').

\end{list}

You should now be able to launch your solver \texttt{baby} on a CUTEr
problem, using the command of Libopt. By typing
\begin{quote}
\texttt{%
\% echo "baby cuter \PROB" | runopt}%
\end{quote}
you should have the same result as by typing ``\texttt{sdbaby} \PROB''.

\subsection{Interfacing a solver with another collection}
\label{s:interfacing-baby-toys}

We consider here the example of the interfacing of the solver
\texttt{baby} with the collection \texttt{toys}, defined in the Libopt
environment in section~\ref{s:hooking-toys-collection}. We recall that
the collection \texttt{toys} is modeled on the Modulopt collection.
More details are given in~\cite{gilbert-2007b}.

\begin{list}{}{\topsep=0.5ex\partopsep=0.5ex\parsep=0.5ex\itemsep=0.5ex
 \settowidth{\labelwidth}{9.}
 \labelsep=0.5em
 \leftmargin=\labelwidth
 \addtolength{\leftmargin}{\labelsep}
}
\item[1.]
Create the directory
\begin{quote}
\texttt{%
\$LIBOPT\_DIR/solvers/baby/toys},
\end{quote}
which will contain the programs/scripts to run the \texttt{baby} solver
on the \texttt{toys} problems, and add the name \texttt{toys} to the
list
\begin{quote}
\texttt{%
\$LIBOPT\_DIR/solvers/baby/collections.lst},
\end{quote}
which indicates that \texttt{baby} can deal with the \texttt{toys}
collection.

\item[2.]
Create the files
\begin{quote}
\texttt{%
\$LIBOPT\_DIR/solvers/baby/toys/all.lst}\\
\texttt{%
\$LIBOPT\_DIR/solvers/baby/toys/default.lst}.
\end{quote}
\begin{list}{{\small$\bullet$}}{\topsep=1.0ex\parsep=0.5ex\itemsep=0.5ex
 \settowidth{\labelwidth}{{\small$\bullet$}}
 \labelsep=0.5em
 \leftmargin=\labelwidth
 \addtolength{\leftmargin}{\labelsep}
 \addtolength{\leftmargin}{\parindent}
 \rightmargin=\parindent
}
\item
The first file (\texttt{all.lst}) must list the problems from the
\texttt{toys} collection that \texttt{baby} is able to solve or, more
precisely, those for which it has been conceived. It can contain {\em
comments}\index{comment (in \texttt{*.lst} files)}, which start with the
`\texttt{\#}' character and go up to the end of the line. The easiest
way of doing this is to start with a copy of the file
\begin{quote}
\texttt{%
\$LIBOPT\_DIR/collections/toys/all.lst},
\end{quote}
which lists all the \texttt{toys} problems, and to remove from the
copied file those problems that do not have the structure expected by
\texttt{baby}. For example, if \texttt{baby} is a solver of
unconstrained optimization problems, remove from the copied file
\texttt{all.lst}, all the problems with constraints. Note that other
lists might exist in the directory \texttt{\$LIBOPT\_DIR/}\abr
\texttt{collections/}\abr \texttt{toys}, which might be more
appropriate to start with than the list \texttt{all.lst}.

\item
The second file above (\texttt{default.lst}) can contain any subset of
the problems listed in the first file (\texttt{all.lst}). This file is
used as the default subcollection when no list is specified in the
\texttt{runopt} script. Therefore, it is often a symbolic link to the
first file \texttt{all.lst}, obtained using the Unix command
\begin{quote}
\texttt{%
ln -s all.lst default.lst}
\end{quote}
in the directory \texttt{\$LIBOPT\_DIR/}\abr \texttt{collections/}\abr
\texttt{toys}.
\end{list}

\item[3.]
Create the executable file
\begin{quote}
\texttt{%
\$LIBOPT\_DIR/solvers/baby/toys/baby\_toys}.
\end{quote}
This is the script launched by \texttt{runopt} to run \texttt{baby} on
a single \texttt{toys} problem. See section~\ref{s:solv-coll-script}
for more information on how to write this script.

\item[4.]
Create the main program
\begin{quote}
\texttt{%
\$LIBOPT\_DIR/solvers/baby/toys/baby\_toys\_main.\textit{src}}.
\end{quote}
This program is very solver dependent and is, with the next step to
which it is linked, the most difficult task to realize. It is the main
program that will be linked with the subroutines describing the problem
from the \texttt{toys} collection selected by the \texttt{runopt}
script, those in the archive \texttt{\textit{prob}.a} (if the selected
problem is \PROB) created by the makefile \texttt{\$LIBOPT\_DIR/}\abr
\texttt{collections/}\abr \texttt{toys/}\abr \texttt{probs/}\abr
\PROB\texttt{/}\abr \texttt{Makefile} in the working directory (see
section~\ref{s:introducing-prob-in-toys}).

If Fortran 90/95 is the adopted language, the easiest way to proceed is
to copy and rename the file
\begin{quote}
\begin{verbatim}
$LIBOPT_DIR/solvers/sqppro/modulopt/sqppro_modulopt_main.f90
\end{verbatim}
\end{quote}
into the file \texttt{\$LIBOPT\_DIR/solvers/baby/toys/}\abr
\texttt{baby\_}\abr \texttt{toys\_}\abr \texttt{main.}\abr
\texttt{f90}. Since this main program is very solver dependent, its
part dealing with the solver will have to be thoroughly modified.
See~\cite{gilbert-2007b} for information on the structure of the
program.

\item[5.]
Create the makefile
\begin{quote}
\texttt{%
\$LIBOPT\_DIR/solvers/baby/toys/Makefile}.
\end{quote}
The aim of this makefile is to tell the Libopt environment how to link
the solver binary with the object files describing the \texttt{toys}
problem selected by the \texttt{runopt}\index{runopt@\texttt{runopt}
(script)} script. If the latter is \PROB, the corresponding object
files will be at link time in the working directory in the archive
\texttt{\textit{prob}.a} (see
section~\ref{s:introducing-prob-in-toys}). The easiest way of doing
this is to start with an existing makefile, like
\begin{quote}
\texttt{%
\$LIBOPT\_DIR/solvers/sqppro/modulopt/Makefile}.
\end{quote}
This one will be copied and renamed into the file
\texttt{\$LIBOPT\_DIR\_DIR/}\abr \texttt{solvers/}\abr
\texttt{baby/}\abr \texttt{toys/}\abr \texttt{Makefile}.
\end{list}

You should now be able to launch the command
\begin{quote}
\texttt{%
echo "baby toys \textit{prob}" | runopt -v}
\index{runopt@\texttt{runopt} (script)}
\end{quote}
where the flag \texttt{-v} (verbose) is used to get detailed comments
from the Libopt scripts, which then tell what they actually do. The
flag \texttt{-t} (test mode) can be used instead, if you want to see
what the scripts would do without asking them to do it.

\section{Final words}

We have presented in this paper the version 1.0 of the Libopt
environment and have motivated its main features. This version of the
software should certainly deserve improvement: we think, for instance,
to the notion of platform, which is presently rather embryonic, to the
possible parametrization of problems, and to the development of a Web
interface with the environment.

Only very few collections and solvers are incorporated in the standard
distribution of this version. Installing new collections and solvers in
a consistent way, maintaining the compatibility with the possible
evolution of the software, may require some expertise and willpower.
Note, indeed, that a change in the software may affect all the cells of
the $\{\mbox{solvers}\} \times \{\mbox{collections}\}$ Cartesian
product. We have the intention to provide these new installations in
the future versions of the software, at least in the optimization
domain, taking profit of the possible contributions of users.

\section*{Acknowledgment}

We thank Jorge Nocedal for having offered the second author hospitality
at Northwestern University (Evanston, USA) when this work was
initiated.

\newcommand{\HOME}{$HOME}

\bibliography{%
\HOME/bibliographies/applications/APPLICATIONS,%
\HOME/bibliographies/optimisation/OPTIM,%
\HOME/bibliographies/informatique/INFO}

\bibliographystyle{plain_a}

\printindex

\end{document}